\documentclass[graybox,natbib,nosecnum]{svmult} 

\usepackage[varvw]{newtx}
\usepackage{amsmath}
\usepackage{array}
\usepackage{makeidx}         
\usepackage{graphicx}        
\usepackage[rightcaption]{sidecap}
\sidecaptionvpos{figure}{c}
\usepackage{multicol}        
\usepackage[bottom]{footmisc}
\usepackage{soul}            
\usepackage[colorlinks,citecolor=blue]{hyperref}

\makeindex             
\def\pe{{\rm P}}
\def\si{{\rm Si}}
\def\es{{\rm S}}
\def\ce{{\rm C}}
\def\en{{\rm N}}
\def\oh{{\rm O}}
\def\be{{\rm Be}}
\def\he{{\rm He}}
\def\li{{\rm Li}}

\def\6v{\sigma v}

\def\mev{~{\rm MeV}}
\def\exp{~{\rm exp}}
\def\ln{~{\rm ln}}
\def\cm{~{\rm cm}}
\def\gcm{~{\rm g~cm}^{-3}}
\def\kmsec{~{{\rm km~sec}^{-1}}}
\def\ergsec{~{\rm erg~sec}^{-1}}
\def\a{\alpha}
\def\b{\beta}
\def\g{\gamma}
\def\G{\Gamma}
\def\d{\delta}
\def\D{\Delta}
\def\e{\epsilon}
\def\vare{\varepsilon}
\def\z{\zeta}
\def\th{\theta}
\def\TH{\Theta}
\def\varth{\vartheta}
\def\i{\iota}
\def\k{\kappa}
\def\l{\lambda}
\def\L{\Lambda}
\def\r{\rho}
\def\varr{\varrho}
\def\s{\sigma}
\def\vars{\varsigma}
\def\t{\tau}
\def\u{\upsilon}
\def\oo{\omega}
\def\O{\Omega}
\def\na{\nabla}
\def\pa{\partial}
\def\p{\prime}
\def\pp{\prime\prime}
\def\sm{~{\rm M}_{\odot}} 
\def\cl{\centerline}
\def\mkt{^{\mu/kT}}

\newbox\grsign \setbox\grsign=\hbox{$>$} \newdimen\grdimen \grdimen=\ht\grsign
\newbox\simlessbox \newbox\simgreatbox
\setbox\simgreatbox=\hbox{\raise.5ex\hbox{$>$}\llap
     {\lower.5ex\hbox{$\sim$}}}\ht1=\grdimen\dp1=0pt
\setbox\simlessbox=\hbox{\raise.5ex\hbox{$<$}\llap
     {\lower.5ex\hbox{$\sim$}}}\ht2=\grdimen\dp2=0pt

\def\boxx{\mathop{\hbox to 0pt{$\sqcup$\hss}\hbox to
0pt{$\sqcap$\hss}\phantom\nabla}}

\begin{document}
\title*{Nuclear Reactions in Evolving Stars (and their theoretical prediction)} 
\author{Friedrich-Karl Thielemann \thanks{corresponding author} and Thomas Rauscher}
\institute{Friedrich-Karl Thielemann \at Department of Physics, University of Basel, Switzerland, and GSI Helmholtz Center for Heavy Ion Research, Darmstadt, Germany \email{f-k.thielemann@unibas.ch}
\and Thomas Rauscher \at Department of Physics, University of Basel, Switzerland, and Centre for Astrophysics Research, University of Hertfordshire, Hatfield, United Kingdom; ORCID:0000-0002-1266-0642
\email{thomas.rauscher@unibas.ch}}
%
%
\maketitle
\abstract{This chapter will go through the important nuclear reactions in stellar evolution and explosions, passing through the individual stellar burning stages and also explosive burning conditions. To follow the changes in the composition of nuclear abundances requires the knowledge of the relevant nuclear reaction rates. For light nuclei (entering in early stellar burning stages) the resonance density is generally quite low and the reactions are determined by individual resonances, which are best obtained from experiments. For intermediate mass and heavy nuclei the level density is typically sufficient to apply statistical model approaches. For this reason, while we discuss all burning stages and explosive burning, focusing on the reactions of importance, we will for light nuclei refer to the chapters by M. Wiescher, deBoer \& Reifarth (Experimental Nuclear Astrophysics) and P. Descouvement (Theoretical Studies of Low-Energy Nuclear Reactions), which display many examples, experimental methods utilized, and theoretical approaches how to predict nuclear reaction rates for light nuclei. For nuclei with sufficiently high level densities we discuss statistical model methods used in present predictions of nuclear reaction cross sections and thermonuclear rates across the nuclear chart, including also the application to nuclei far from stability and fission modes.}

\section{Describing nuclear composition changes via nuclear reaction rates}
\label{sec:nuc_reac}

Before going into more details about the nuclear reactions in evolving stars in the next sections, we present here a short overview of the tools used to calculate abundances changes via nucleosynthesis networks and the relevant reaction rates. 
Abundances of nuclei can defined by their mass fraction ($X_i$), i.e. the percentage of their mass in species $i$ (or density $\rho_i$) with respect to the total mass (density) in two possible ways:
\begin{subequations}
\begin{alignat}{3}
\label{eq:abundances1}
    X_i&=\frac{\rho_i}{\rho}=\frac{n_im_i}{\rho}=\frac{n_i\mathcal{A}_im_u}{\rho}=\frac{n_i}{(\rho/m_u)} \mathcal{A}_i\\
\label{eq:abundances2}
    X_i&=\frac{n_i m_i}{\rho N_\mathrm{A}} N_\mathrm{A} =\frac{n_i}{\rho N_\mathrm{A}}\mathcal{A}_im_u N_\mathrm{A}=\frac{n_i}{\rho N_\mathrm{A}}\mathcal{A}_iM_u,
\end{alignat}    
\label{eq:abundances}
\end{subequations}
with $n_i$ being the number density, $\rho$ the overall mass density, $N_\mathrm{A}$ the Avogadro number, $m_u=\frac{1}{12}m(^{12}$C) the nuclear mass unit, $\mathcal{A}_i$ the relative atomic mass, and $M_u=N_\mathrm{A}m_u$ the molar mass constant. Replacing the relative atomic mass $\mathcal{A}_i$ by $A_i=N_i+Z_i$, i.e. the nuclear mass number given by the sum of neutrons and protons in a nucleus, introduces an error in the per-mille range due to the nuclear binding, which is usually neglected. The abundance of a nucleus, rather than its mass fraction, should not reflect its weight and is defined as $Y_i=X_i/\mathcal{A}_i\approx X_i/A_i$, which can be expressed, utilizing Eq.(\ref{eq:abundances}), by
    $Y_\mathrm{i}={n_\mathrm{i}}/({\rho N_\mathrm{A}})M_u ={n_\mathrm{i}}/{(\rho/m_u)}$.
Until 2019 $M_u$ had the value of $10^{-3}$~kg/mole or 1~g/mole in cgs units. This led \cite{Fowler.Caughlan.Zimmerman:1967} and the following literature to omit this factor of 1 in the definition of the abundance (but also introducing the dimension of g/mole). According to the latest CODATA evalution \citep[][Table XXXI]{Tiesinga.ea:2021} this is still close to being correct (numberwise) with a precision of $3\times 10^{-10}$, but introduces somewhat confusing dimensions for the abundances. We will continue to utilize here the traditional expressions, but want to make clear that, when changing all terms $\rho N_\mathrm{A}$ - appearing in equations introduced here and later in this text - to $\rho/m_u$, like in $Y_\mathrm{i}={n_\mathrm{i}}/{(\rho/ m_u})$ \citep[as often done in more modern approaches, see e.g.][]{Cowan.Sneden.ea:2021}, the expressions stay valid and lead for the mass fractions $X_i=A_i Y_i$ as well as the abundances $Y_i$ to dimensionless numbers. 
 All mass fractions add up to unity:
 \begin{equation}
     \sum_i X_i = \sum_iA_i Y_i = 1.
 \end{equation}
Astrophysical environments carry generally no charge, i.e. the total number of protons has to be equal to the number of electrons 
\begin{equation}
    \sum_i Z_i Y_i = Y_\mathrm{e}=\frac{n_\mathrm{e}}{\rho N_\mathrm{A}},
\end{equation}
with the electron fraction ($Y_\mathrm{e}$) being on the one hand defined by the electron number density but also equal to the number of protons per nucleon (i.e., protons plus neutrons) $Y_e={{\sum_i Z_i Y_i}/{\sum_i A_i Y_i}}$.

The individual abundances can be calculated with a nuclear reaction network 
based essentially on $r$, the number of 
reactions per volume and time that can be expressed, when targets and projectiles follow specific distributions $dn$, by
\begin{equation}
r_{ij}=\int \sigma\cdot\vert\vec v_i-\vec v_j\vert d n_i d n_j.
\label{eq:reac-basic}
\end{equation}
The evaluation of this integral depends on the type of particles and
distributions which are involved. For nuclei $i$ and $j$ in an astrophysical 
plasma, obeying a Maxwell--Boltzmann distribution 
we find $r_{ij}=n_in_j \langle \sigma v \rangle_{ij}$
where $\langle \sigma v \rangle$ is integrated over the relative bombarding energy and is only a function of temperature $T$
\begin{equation}
    \langle \sigma v\rangle_{ij}(T) = \left( \frac{8}{\pi \mu_{ij}} \right)^{1/2} (kT)^{-3/2} \int_0^\infty E \sigma(E) \exp \left( -\frac{E}{kT} \right) \mathrm{d}E,
    \label{sigmav}
\end{equation}
where $\mu_{ij}$ is the reduced mass of targt and projective $i$ and $j$.
For a reaction with photons we have $j=\gamma$, i.e.
in this case the projectile $j$ is a photon. The relative velocity
is the speed of light $c$, the distribution $dn_j$ is the Planck
distribution of photons. As the relative velocity
between the nucleus and the photon is a constant ($c$), and the
photodisintegration cross section is only dependent on the photon energy $E_\gamma$, the integration over $dn_i$ can be easily performed, resulting in

\begin{align}
r_{i\gamma}&=n_i\frac{1}{\pi^2c^2\hbar^3}
\int{\frac{\sigma_i(\gamma;E_\gamma)E^2_\gamma}{\exp(E_\gamma/kT)-1}dE_\gamma} = n_i\lambda_{i\gamma}(T)\nonumber \\
\lambda_{i\gamma}(T)&=\frac{1}{\pi^2c^2\hbar^3}
\int{\frac{\sigma_i(\gamma;E_\gamma)E^2_\gamma}{\exp(E_\gamma/kT)-1}dE_\gamma}.
\label{eq:reac-photo}
\end{align}
Contrary to the reactions among nuclei or nucleons, where both reaction
partners are following a Boltzmann distribution, this expression has only a linear dependence on number densities. The integral
acts like an effective (temperature dependent) decay constant of nucleus $i$. 
Electron captures behave in a similar way as the mass difference between nucleons/nuclei and electrons is huge and the relative velocity is with high precision given by the electron velocity. This leads to

\begin{equation}
r_{i;e}=n_i\int\sigma_e(v_e)v_edn_e
=\lambda_{i;e}(\rho Y_e,T) n_i.
\label{eq:reac-electron}
\end{equation}
This is an expression similar to that for photodisintegrations, but now we have a
temperature and density dependent ``decay constant'', because the electron distribution has also a density dependence and can change to a degenerate Fermi gas (for details see the chapter by Suzuki and \cite{Langanke.Martinez:2021}).
In principle neutrino reactions with nuclei would follow the same line, because the neutrinos, propagating essentially with light speed, would lead to a simple integration over the neutrino energies. However, as neutrinos, interacting very weakly, do not necessarily obey a thermal distribution for local conditions, their spectra depend on detailed transport calculations
(see the chapters by Rrapaj \& Reddy, Fuller \& Grohs, Wang \& Surman, Famiano et al., Fr\"ohlich, Wanajo, and Obergaulinger), leading to 

\begin{equation}
r_{i\nu}=n_i\int\sigma_e(E_{\nu})cdn_{\nu}(E_{\nu})
=\lambda_{i\nu}({\rm transport}) n_i.
\label{eq:reac.neutrino}
\end{equation}
Finally, for normal decays, like beta- or alpha-decays or ground-state fission with a half-life 
$\tau_{1/2}$, we obtain a similar equation with 
a decay constant $\lambda_i$=$ln2/\tau_{1/2}$ (see the chapter by Rubio et al., Block et al., Oberstedt \& Oberstedt, Schunk, and Kowal \& Skalski) and

\begin{equation}
r_i=\lambda_i n_i. 
\label{eq:reac-decay}
\end{equation}
In this case the change of the number density due to decay is 
$\dot n_i=-\lambda_in_i$, with the solution $n_i=n_i(0)e^{-\lambda_it}$ and 
$n_i(\tau_{1/2})=\frac{1}{2}n_i(0)$. The decay half-life of a nuclear
ground state is a constant. 

Adding all these different kind of reactions, we can describe the time derivative of abundances $Y=n/\rho N_A$ as the difference of production and destruction terms 
with a differential equation for each species.
Both the production and destruction channels include particle-induced reactions, decays, photodissociation, electron capture, etc. For every nucleus $i$ the abundance is given by a differential equation
\begin{align}
    \dot{Y}_i = &\sum_jN^i_j\lambda_jY_j+ \sum_{j,k} \frac{N_{j,k}^i}{1+\delta_{jk}}\rho N_\mathrm{A} \langle \sigma v\rangle_{j,k} Y_j Y_k + \notag \\
    &\sum_{j,k,l} \frac{N_{j,k,l}^i}{1+\Delta_{jkl}}\rho^2 N_\mathrm{A}^2 \langle \sigma v\rangle_{j,k,l} Y_j Y_k Y_l,
\label{eq:networkequ}
\end{align}
where the factors $1/(1+\delta_{jk})$ and $1/(1+\Delta_{jkl})$ prevent double counting of reactions in two- and three-body reactions, respectively. $\Delta_{jkl}$ has the value 0, 1 or 5, so that, dependent on the multiplicity of identical partners, the denominators are equal to 1!, 2!, or 3!.
For methods to solve these (often huge) stiff sets of ordinary differential equations see e.g. \cite{Hix.Thielemann:1999b} and \cite{timmes99}, a general overview of utilized codes in astrophysical applications as well as of publicly available tools is given in \cite{Reichert.ea:2023}.

Knowing the cross section $\sigma(E)$ of a nuclear reaction, $\left< \6v
\right>$ can easily be determined, provided that the participating
nuclei obey Boltzmann statistics. In the following section we
want to utilize this method to predict abundance/composition changes. We will first go through the stellar burning stages identifying the
important reactions, while we rely on the sections by M. Wiescher and P. Descouvement how they are determined experimentally or theoretically for light nuclei, before introducing statistical model methods for intermediate mass and heavy nuclei.

One extreme case should however be mentioned, which occurs at sufficiently high temperatures when on the one hand the Coulomb barriers can be overcome for capture reactions and on the other hand the reverse photodisintegrations are efficient due to the high energy photons of the corresponding Planck distribution. For a reaction $i(p,\gamma)m$, with $i$ standing for nucleus $(Z,A)$ and $m$ for $(Z+1,A+1)$, the relation between the forward rate $\left<\sigma v\right>_{i;p,\gamma}$  and the photodisintegration rate $\lambda_{m;\gamma,p}$ is given by
\begin{equation}
\lambda_{m;\gamma,p}(T)=\frac{g_p G_i}{G_m}
\left(\frac{\mu_{in} kT}{2\pi\hbar^2}\right)^{3/2}
\exp(-Q_{i;p,\gamma}/kT)\left<\sigma v\right>_{i;p,\gamma}, 
\label{eq:1}
\end{equation}
containing the reduced mass $\mu_{in}$ and the capture Q-value for the reaction, $g_p=2\times(1/2)+1$ for the protons and the partition functions $G$ for the nuclei involved.
When utilizing that the difference between forward proton capture and backward photodisintegration flux for nucleus $i$, i.e.
$\dot Y_i=-\rho N_A \left<\sigma v\right>_{i;p,\gamma}Y_pY_i+\lambda_{m;\gamma,p}Y_m$ vanishes in chemical equilibrium. Making also use of the relation between  $\left<\sigma v\right>_{i;p,\gamma}$ and $\lambda_{m;\gamma,p}$, based on detailed balance, results in 
\begin{equation}
{{Y_m}\over {Y_i}} =\rho N_A Y_p {{G_m}\over{2G_i}}
                            \left[{{A_m}\over{A}_i}\right]^{3/2}
                        \left[{{2\pi\hbar^2}\over{m_ukT}}\right]^{3/2}         \exp\left({{Q_{i;p,\gamma}}\over{kT}}\right).
\label{eq:2}
\end{equation}
Such a chemical equilibrium for a specific forward and reverse reaction can also apply for e.g. neutron or even alpha captures. When temperatures are sufficiently high to permit a whole sequence of proton and neutron captures up to nucleus $(Z,A)$ or $(Z,N)$, a complete chemical equilibrium (also termed nuclear statistical equilibrium NSE) can be attained, linking neighboring nuclei via neutron or proton captures as in Eq.(\ref{eq:2}) up to
nucleus $(Z,N)$. The approach to such NSE conditions in discussed in the following section on burning phases in stellar evolution.

It is important to note that the reaction cross sections appearing in Eqs.\ (\ref{eq:reac-basic}) -- (\ref{eq:reac.neutrino}), as well as the reactivities $\left<\sigma v\right>$ in Eq.\ (\ref{eq:networkequ}) must account for the thermal
excitation of nuclei in an astrophysical plasma and may differ from laboratory cross sections, determining cross sections of target nuclei in their ground states only. In an astrophysical environment, nuclei are found in excited states with a probability given by the plasma temperature $T$ and the excitation energies $E_i$ of low-lying excited states \citep{Ward.Fowler:1980}. Thus, the actual \textit{stellar} reaction rate $r^*$ is given by a weighted sum of
rates of reactions on individual excited states bombarded by a thermal (or non-thermal, in case of neutrinos) distribution of projectiles \citep{Rauscher:2020},
\begin{equation}
r^*=P_0 r_0 + P_1 r_1 + P_2 r_2 + \dots\quad,
\label{eq:thermalrate}
\end{equation}
with the individual population factor of an excited state
given by
\begin{equation}
P_i = \frac{2J_i+1}{G} \mathrm{e}^{-E_i/(k T)}\quad.\label{eq:pop}
\end{equation}
Here, $G$ is the nuclear partition function of the target nucleus and $J_i$ the spin
of excited state $i$ ($i=0$ for the ground state). 
The reciprocity
relation shown in Eq.\ (\ref{eq:1}) also only holds when such stellar rates are used.
Instead of summing over individual contributions from several rate integrals as shown in Eq.\ (\ref{eq:thermalrate}), the form of a single integral (as in Eq.\ \ref{eq:reac-basic}) can be retained by using a stellar cross section $\sigma^*$ as
defined in Eq.\ (\ref{eq:HF*}) \citep{Holmes.Woosley.ea:1976,Rauscher:2020,Rauscher:2022}.

The relative contribution to the stellar rate of reactions on nuclides in the ground state can be
computed from
\begin{equation}
X_0^*(T)=\frac{r_0}{r^*G}\,,
\end{equation}
where $r_0$ is the reaction rate calculated from reactions on the ground state only.
Since the population of excited states depends on the excitation energies of available
excited states, the impact of excited states is smaller (and thus $X_0\approx 1$) for light nuclides with large level spacings
but is non-negligible ($X_0<1$) for intermediate and heavy nuclides even at such low temperatures $k_\mathrm{b}T$
of a few keV, as encountered, e.g., in the s-process \citep{Rauscher:2022}.
In hot,
explosive astrophysical environments $X_0$ becomes tiny and most contributions to the
stellar rate come from reactions on excited states.

\section {Stellar burning stages}

Stellar burning stages are characterized by reactions involving
increasingly larger nuclei (and charges). The necessary velocities
(energies) to overcome the Coulomb barriers of heavier nuclei increase
with charge $Z$ and are related to the environment temperature.
When one type of reaction sequence is started, 
the energy loss of a star (its luminosity) is equal to
the total energy generation, once stable burning conditions have been
attained. The burning temperature stays approximately
constant until the fuel is exhausted.  At that point decreasing pressure
support will lead to a contraction, which again causes  a 
temperature increase, due to the release of gravitational binding energy. At
some critical temperature threshold the nuclei of the
next burning ``fuel'' can overcome the Coulomb barriers in the main
reactions of this next burning stage.  In massive stars the sequence of
burning stages is given by the fuels $^1$H, $^4$He, $^{12}$C, $^{20}$Ne,
$^{16}$O and $^{28}$Si (see also the chapter by K. Nomoto on the impact of reaction rate uncertainties on stellar evolution). We want to discuss the details and the main reaction
sequences of all these burning stages in the following subsections.

\subsection {Hydrogen Burning}
\subsubsection{The PP-Cycles}
\begin{table}[h!]
\centering
\caption{The PP-Cycles in Hydrogen Burning}
\begin{tabular}{c r l r}
\hline
\hline
 cycle& reaction & $\tau$(years) & Q(MeV) \\
\hline
PPI  & $^1$H(p,$e^+\nu$)$^2$H &7.9$\times 10^9$&0.420\\
(pep reaction)     & $^1$H(p$e^-,\nu$)$^2$H &3.7$\times 10^{12}$&1.442\\ 
     & $^2$H(p,$\g$)$^3$He &5.9$\times 10^{-8}$&5.493\\
     & $^3$He($^3$He,2p)$^4$He &1.4/$Y_3$&12.859\\
PPII & $^3$He($\a,\g$)$^7$Be &1.1$\times 10^6$&1.586\\
     & $^7$Be($e^-,\nu$)$^7$Li &2.9$\times 10^{-1}$&0.861\\
     & $^7$Li(p,$\a$)$^4$He &4.3$\times 10^{-5}$&17.347\\
PPIII& $^7$Be(p,$\g$)$^8$B &1.8$\times 10^2$&0.135\\
     & $^8$B($e^+\nu$)$^8$Be$^*(\a)~^4$He &3.5$\times 10^{-8}$&18.078\\
PPIV (Hep reaction) & $^3$He(p,$e^+\nu$)$^4$He& 3.7 $\times 10^{7}$ &19.795\\ 
\hline
\end{tabular}
\label{tab:6.pp}
\end{table}


In a pure hydrogen gas the reaction sequences in Table \ref{tab:6.pp} occur at $T\approx
10^7$ K. We give the individual sub-cycle name, the Q-value and a
lifetime for each reaction. Although the purely nuclear Q-value of the
first pp-reaction is only 0.420$\mev$, the effective Q-value is also
1.442$\mev$ as in the pep-reaction, because the positron annihilates with an 
electron in the plasma and produces 1.022$\mev$ in photons.
In the case of a decay, lifetimes are related to
the decay constant $\lambda$ via $\tau=1/\lambda$, or half-life with
$\lambda=\ln(2)/\tau_{1/2}$. After a lifetime $\tau$ passed in a reservoir
of unstable nuclei, a fraction $1/e$ will remain. For reactions involving 
a projectile and target, the target lifetime $\tau_i$ is given by
$\dot Y_i=-{1\over\tau_i}Y_i$, containing the reaction rate (i.e. requiring a knowledge of density and temperature) and the projectile abundance. 
The table values were obtained assuming
$\rho=100\gcm$, and
$T=1.5\times 10^7{\rm K}$, based on reaction rate evaluations of \cite{Angulo.ea:1999} and \cite{Adelberger.ea:2011}. The abundances are taken from an intermediate state of H-burning with $Y_{\rm H}=0.5,~Y_{\rm He}={0.5/4}$ (i.e. half of the H is burned to He), while in a steady flow equilibrium the abundances of the intermediate nuclei are close to negligible. For most of the reactions in Table \ref{tab:6.pp} the projectile is either $^1$H or $^4$He, i.e. the lifetimes can be determined with these known abundances, an exception is  $^3$He($^3$He,2p)$^4$He. When taking typical steady flow values of the order $Y_3=Y_{\rm ^3He}\approx 10^{-4}$, this leads to a lifetime of about $10^4$ years.

In all cases (i.e. PPI, PPII, PPIII or PPIV) 
the net reaction is the fusion of 4 protons into one alpha
particle ($^4$He) with two $\b$-decays or electron captures among the
intermediate reactions. In the PPI-cycle six hydrogen nuclei (protons) are 
required in total, two are finally released in the $^3$He-$^3$He-reaction,
together with one helium nucleus, i.e. the net reaction is 4 $^1$H$\rightarrow
^4$He. The slowest reaction (of relevance) in all the cycles is the first one listed.  The
pep-reaction reaction is the slowest of all, because a three body collision is a very
rare event. It is, however, unimportant for H fusion, as the pp-reaction burns and
produces the same nuclei, only the properties of the emitted neutrinos
are different.

The PPIV cycle or Hep reaction is mostly important for relatively high energy neutrinos, but, as can be seen from the lifetime of the Hep reaction in comparison to $^3$He(p,$\gamma)$, it carries also about one third of the latter PPII flux from $^3$He to $^4$He.
The pp-reaction is very slow, because a diproton ($^2$He) is
unstable and breaks up immediately. 
during the very limited time of the existence of a diproton 
a weak transition of one of the protons into a neutron occurs,
The production of deuterium ($^2$H) is only possible
during the very limited time of the existence of a diproton, 
when a weak transition of one of the protons into a neutron occurs.
Due to its rarity, 
this reaction rate is based entirely on theoretical predictions.
A cross section that small is not accessible by experiments. It should,
nevertheless, be quite accurate, because only the fundamental constants
of weak interaction theory enter. Thus a description
of nuclear many body systems (nuclei), which may or may not be correct, is not required 
\citep[see the detailed discussion in e.g.][]{rolfs.rodney:1988,Bahcall:1989,Adelberger.ea:2011}.

The other pp-cycle reactions are studied by experiments \citep[see for historical reasons the detailed discussions in][]{rolfs.rodney:1988}, \cite{Caughlan.Fowler:1988} and updates in \cite{Angulo.ea:1999} and \cite{Adelberger.ea:2011}. Typically, cross-sections are measured at somewhat higher energies, where cross sections are larger and then extrapolated
to the solar energies of interest \citep[see e.g.][]{Formicola.ea:2016}.
There existed some uncertainty in the reaction
$^7$Be(p,$\gamma)^8$B, but recent evaluations by \cite{Tursunov:2021} provide nice agreement with the \cite{Adelberger.ea:2011} rate. $^7$Be is unstable against electron capture (EC),
not $\beta^+$-decay, due to a small Q-value. The terrestrial lifetime,
$\tau$=77d
(half life $\tau_{1/2}$=53d), is caused by capture of a bound K-shell
electron. For conditions in the center of stars, where atoms are
completely ionized, electron capture has to occur with electrons from
the continuum electron distribution, which leads to a lifetime of 106d
for the conditions of Table \ref{tab:6.pp} \citep{Adelberger.ea:2011}. The terrestrial lifetime is long enough to
produce radioactive $^7$Be targets and several direct measurements of the
$^7$Be(p,$\gamma)^8$B reaction have been performed by radioactive ion beam facilities \citep[see e.g.][]{Adelberger.ea:2011}. In such a case $^7$Be, produced by a primary
nuclear reaction, is accelerated in a second step onto an existing
hydrogen gas target. Thus, the role of target and projectile is
reversed and the reactions.

After a burning cycle proceeds for a time scale in excess of 
all reaction time scales in the cycle, except for the longest
one, an abundance equilibrium is attained.
In general such a situation is called a steady flow equilibrium, where
the slowest link determines the speed and all other reactions adjust to
that reaction flow.
The destruction of $^1$H (and the build-up of $^4$He) is caused, in
temperature and density dependent fractions,
by the four PP cycles, determined by the branch points at
$^3$He and $^7$Be, which open the PPII, PPIV and 
PPIII-cycles respectively. 
Only the latter branching is temperature and density
dependent, because a decay [$^7$Be($e^-,\nu)^7$Li] and a capture reaction
have a different density dependence.
Knowing the hydrogen abundance at a given time (and the density and
temperature for determining the reaction rate) automatically determines
the steady flow abundances of $^2$H and $^3$He. From Table \ref{tab:6.pp} and the destruction time scales we
see that only A=1 and 4 can have non-negligible abundances, more than four orders of magnitude larger than those of any other cycle member. 
Mass conservation gives $Y_4=(1-Y_1)/4$. As can be seen from Table \ref{tab:6.pp}, the much larger lifetime of $^3$He in the Hep reaction by 2-3 orders of magnitude make the PPIV cycle contribution unimportant for energy generation, but this sub-cycle can contribute the neutrinos with the highest energies.

\subsubsection{The CNO-Cycles}
\label{sec:6.1.2}

If heavier elements (CNO) are already present in the stellar plasma, another
chain of reactions is possible, when temperatures are large enough to permit
the penetration of the larger Coulomb barriers. These reactions can 
be faster at appropriate temperatures than the weak pp-reaction which 
involves a p($e^+\nu)$n transition. Such a sequence of reactions,
involving the most abundant lightest nuclei beyond $^4$He (C, N, and O),
are known as the CNO-cycles.

\begin{table}[h!]
\centering
\caption{The CNO-Cycles in Hydrogen Burning}
\begin{tabular}{r c}
\hline
\hline
\ cycle & reaction sequence \\
\hline
CNOI &
$^{12}$C(p,$\gamma)^{13}$N($e^+\nu)^{13}$C(p,$\gamma)^{14}$N
(p,$\gamma)^{15}$O($e^+\nu)^{15}$N(p,$\alpha)^{12}$C \\
CNOII & $^{15}$N(p,$\gamma)^{16}$O(p,$\gamma)^{17}$F
($e^+\nu)^{17}$O(p,$\alpha)^{14}$N \\
CNOIII & $^{17}$O(p,$\gamma)^{18}$F($e^+\nu)^{18}$O(p,$\alpha)^{15}$N \\
CNOIV & $^{18}$O(p,$\gamma)^{19}$F(p,$\alpha)^{16}$O \\
\hline
\end{tabular}
\label{tab:6.CNO}
\end{table}
Most of the reactions listed in Table \ref{tab:6.CNO} are also studied experimentally, but again the extrapolation to low energies in the keV range is important to understand the behavior at Gamow peak energies \citep[see e.g.][]{Adelberger.ea:2011}. A major uncertainty has been the 
$^{17}$O(p,$\alpha)$/$^{17}$O(p,$\gamma)$-branching between the CNOII and
CNOIII cycle, related mostly to a low-lying
65keV resonance \citep[see e.g.][]{rolfs.rodney:1988} which has recently been determined by \cite{Bruno.ea:2016} in a direct measurement.
The CNOIV-chain is not of great importance at all temperatures, the
flux is more than a factor 100 smaller than in the CNOIII-chain.
At lower temperatures we have only one single
chain without branchings, which is the CNOI chain.  
Therefore, the flux passing through each reaction in case of a steady flow
equilibrium, $C=\rho N_A\left<i1\right>Y_iY_1$, must be identical. 
$^{14}$N(p,$\g)^{15}$O has the smallest reaction rate, which means that
the larger rates for other reactions have to be balanced with smaller
abundances and
almost negligible abundances of the other CNO nuclei remain. In this case, when assuming
initial solar abundances for all CNO-nuclei, 
a steady-flow equilibrium assembles all CNO nuclei in $^{14}$N. This leads to
\begin{equation}
Y_{14}\approx {X_{CNO}\over 14},\label{eq:6.5}
\end{equation}
Solar system abundance determinations have changed in recent years. While meteoritic measurements help to determine the relative abundances of non-volatile elements, when requiring the ratio to H and He also solar spectra have to be analyzed. Early 1D spherically symmetric approaches of the solar photosphere have been replaced by multi-dimensional 3D analyses \citep{Anders.Grevesse:1989,Asplund.Grevesse.ea:2009}.
This led to a change for $X_{CNO}$, i.e. the mass fraction of all CNO nuclei in a solar composition, from $1.4\times 10^{-2}$ to $9.33\times 10^{-3}$ \citep[which is the mass fraction
of all CNO nuclei in a solar composition with 
$Y_C=2.18\times 10^{-4}$, $Y_N=5.73\times 10^{-5}$, $Y_O=3.7\times 10^{-4}$ and $X_{CNO}=12Y_C+14Y_N+16Y_O$, see e.g.][]{Amarsi.Grevesse.ea:2021}. 

For solar CNO-abundances
($Y_{14}=9.33\times 10^{-3}/14$) and $Y_1=0.5$, the CNO-cycles dominate over
the PP-cycles beyond $T=1.75\times 10^7$K.
Under solar conditions the PP-chains dominate by more than one order of
magnitude.
As in the PP-cycles the final product of H-burning is $^{4}$He.
Essentially all initial CNO abundances are transferred to $^{14}$N, with none
being burned to heavier nuclei. It is important to note that 
the CNO nuclei only act as a catalysts.

A cycle similar to the CNO cycle, requiring slightly higher temperatures is the NeNaMg-cycle, shown in Table \ref{tab:6.NaNaMg}.
\begin{table}[h!]
\centering
\caption{The NeNaMg-Cycles in Hydrogen Burning}
\begin{tabular}{r c}
\hline
\hline
\ cycle & reaction sequence \\
\hline
NeNaMgI &
$^{20}$Ne(p,$\gamma)^{21}$Na($e^+\nu)^{21}$Ne(p,$\gamma)^{22}$Na
(p,$\gamma)^{23}$Mg($e^+\nu)^{23}$Na(p,$\alpha)^{20}$Ne \\
NeNaMgII & $^{21}$Na(p,$\gamma)^{22}$Mg($e^+\nu)^{22}$Na \\
\hline
\end{tabular}
\label{tab:6.NaNaMg}
\end{table}

For updates beyond the compilations mentioned above \citep{Angulo.ea:1999,Adelberger.ea:2011} see the impressive recent activities in underground labs which attempt to measure the related cross sections down to the relevant stellar energies:
i.e. ongoing investigations in LUNA\footnote{\url{https://luna.lngs.infn.it/index.php/new-about-us}}, CASPAR
\footnote{\url{https://caspar.nd.edu/}} and JUNA \citep{JUNA-Liu:2022}, which avoid background noise and permit cross section measurements down to energies in the 50\,keV region. This also applies for reactions of interest in the upcoming sections of He- and C-burning.

\bigskip
\subsection {Helium Burning}
\medskip
\subsubsection {Main Reactions}
\label{sec:6.2.1}

There is no stable $A=5$ nucleus. Proton or neutron captures on $^{4}$He
cannot produce heavier nuclei in stellar environments. This is the
reason why hydrogen burning does not proceed beyond $^4$He. The ashes of
hydrogen burning consist almost entirely of $^4$He, except for minor
amounts of $^2$H, $^3$He, $^7$Li, and pre-existing heavy elements.
$^3{\rm He}(^4{\rm He},\g)^7{\rm Be}$ produces $^7\be$ and, after decay, 
$^7\li$. In hydrogen burning Li is destroyed again via
$^7$Li($p,\alpha)^4$He; in helium burning the small abundance of $^3\he$ 
left over from hydrogen burning is exhausted very fast.
The only reaction possible among $^4$He-nuclei is 

\bigskip
\cl{$^4$He+$^4$He$\rightleftharpoons ^8$Be,}
\bigskip

\noindent
which produces the unstable $^8$Be nucleus, decaying on a timescale of 
2.6$\times 10^{-16}$s via $\a$-emission. 
However, increasing temperatures and densities lead to a large production
of $^8\be$, which decays immediately, but a tiny abundance remains in
(chemical) equilibrium (where forward and backward reaction cancel), leading to (with 4 and 8 standing for
$^4$He and $^8$Be

\begin{equation}
{\dot Y_8={1\over 2}\rho N_A\left<4,4\right>Y^2_4-\l_8Y_8=0\ \ \ \ 
\l_8={1\over\tau_8}~~~~\G_8\tau_8=\hbar~~~~\l_8={\G_8\over\hbar}}.
\label{eq:6.9}
\end{equation}

The life time of $^8$Be is related to the width of the ground state via
the Heisenberg uncertainty principle. The equilibrium abundance of
$^8$Be is consequently

$$ Y_8={\hbar\over 2\G_8}\rho N_A\left<4,4\right>Y^2_4.$$

\noindent Another alpha-capture can follow, leading to the production of $^{12}$C
\begin{align}
\dot Y_{12}&=\rho N_A\left<4,8\right>Y_4Y_8\nonumber \\
&={\hbar\over 2\G_8}\rho^2 N^2_A\left<4,4\right>\left<4,8\right>Y^3_4\label{eq:6.10}\\
&\equiv{1\over 3!}\rho^2N^2_A\left<4,4,4\right>Y^3_4.\nonumber
\end{align}
The last line is the definition of the triple-alpha-rate, treating it
as if it were a real three-body reaction. In the case of 
a collision of
three identical particles, we would  have a double counting
of possible reactions by taking the full helium abundance. The
overcount is given by the permutations of the three particles, which would
all end up in the same final $^{12}$C nucleus, whatever the 
reaction sequence is.
The number of permutations among three particles is 3!, and
thus is included in the above definition. 
This
leads to the triple-alpha-reaction rate

$$^4\he(2\a,\g)^{12}{\rm C}~~~~Q=6.445\mev$$
\begin{equation}
N^2_A\left<4,4,4\right>=N^2_A{3\hbar\over\G_8}\left<4,4\right>
\left<4,8\right>. 
\label{eq:6.11}
\end{equation}

A new feature is encountered in the triple-alpha reaction sequence which does not
occur in regular two-dody reactions. The second alpha capture proceeds on a
target in an unstable state with a natural width. If $^8$Be was not
produced at the resonance energy, but in the low energy tail, the
energetics of the second alpha capture (with target $^8$Be) 
would be changed
with respect to resonances in $^{12}$C. An appropriate approach (especially of
importance at relatively low temperatures and high densities) which is also important in other 
applications, like explosive H-burning, has been outlined in a number of 
publications \citep[see e.g.][]{Nomoto.Thielemann.Miyaji:1985,Goerres.Wiescher.Thielemann:1995}.

A possible alpha capture reaction on the product $^{12}$C depends on $T$,
with
$$^{12}\ce(\a,\g)^{16}\oh~~~~Q=7.161\mev.$$
$^{16}\oh(\a,\g)^{20}$Ne is blocked by a very small cross section, due
to missing resonances at the appropriate energies in $^{20}$Ne. 
Thus, the dominant ash of hydrogen burning ($^4$He) is burned in helium
burning to $^{12}$C and $^{16}$O.

For typical burning conditions of $T=(1-2)\times 10^8{\rm K}$ and
$\rho=3\times 10^2-10^4\gcm$ the result is a mixture of $^{12}$C and
$^{16}$O, with $^{16}$O dominant at higher temperatures. This
situation is, however, complicated by additional aspects of stellar
evolution and overshadowed by a large experimental uncertainty
of the $^{12}$C($\alpha,\g)^{16}$O cross section \citep[see e.g.][and the footnotes at the end of the H-burning section]{DeBoer.ea:2017}.
The effectiveness of the  $^{12}$C($\alpha,\gamma$)$^{16}$O reaction
depends on the temperature of He-burning, i.e. the stellar mass
{\it and} the cross section of the reaction itself. Since  
the temperature increases during He-burning, it is also important
to note that the mixing of fresh He into the burning core, at 
the end of the core 
burning phase, can mimic a high reaction rate. 
Thus, the astrophysical treatment of convection is also important 
in this case. 
All of these effects contribute
to the ``effective'' transformation of $^{12}$C into $^{16}$O. The
C/O-ratio at the end of He-burning determines the further evolution
of the star. Small carbon concentrations, or completely depleted carbon,
can  result in the case that either of the two burning stages of C-burning and Ne-burning 
is not occurring \citep[for a detailed discussion see e.g.,][or the chapter by K. Nomoto]{ojima18}.

\subsubsection{Neutron Production in Helium Burning}
\label{sec:6.2.2}

Essentially all material in CNO-nuclei, present in the initial stellar 
composition, is processed to $^{14}$N during hydrogen burning.
This leads to a side branch of helium burning
$$^{14}\en(\a,\g)~^{18}{\rm F}(e^+\nu)~^{18}\oh(\a,\g)~^{22}{\rm Ne}.$$

\par\noindent
This is the one reaction in helium burning, where protons are changed
to neutrons, leading to a
decrease of $Y_e$. For a solar initial composition with 1.4\% by mass in
CNO nuclei (which turned into $^{14}$N in hydrogen burning), an abundance
of $Y(^{22}$Ne$)=Y(^{14}$N$)=0.014/14=1.0\times 10^{-3}$ is obtained.
This non-symmetric nucleus is characterized by $N-Z=2$ and results in $Y_e=0.499$.

Some of the $^{22}$Ne undergoes another alpha-capture to produce
$^{26}$Mg. A smaller fraction (dependent on temperature) follows an
endoenergetic reaction $^{22}$Ne($\a,n)^{25}$Mg  ($Q=-0.482\mev$), i.e.
only alpha energies (center of mass) beyond 0.482$\mev$ can induce this
reaction. Such energies can be attained for $T>3.5\times 10^8$K. This is larger than the
temperatures in helium burning discussed above. On the other hand, the
energy threshold does not have to be at the center of the Gamow window
of Eq.(\ref{sigmav}). The high energy tail
also gives a contribution. It is clear, however, that this neutron
producing reaction is only active in late phases of core helium burning,
which experience high temperatures. 

When helium burning occurs in burning shells and shell flashes (to be 
discussed later), additional features can occur. High
temperatures can be attained, and in some cases hydrogen is mixed into the 
helium burning zones, as a result of thermal pulses (or helium shell flashes).
This permits proton captures on $^{12}$C, which produce $^{13}$N, decaying
to $^{13}$C. The latter is also a very efficient neutron source as 
due to a strong
$(\a$,n)-reaction. Both neutron sources ($^{22}$Ne and $^{13}$C)
lead to neutron production and the build-up of heavy elements via 
neutron capture and $\b^-$-decay of unstable nuclei. If neutron capture
time scales are longer than beta-decay time scales, this
is called a slow neutron capture process (s-process), which will be
discussed later in more detail. The important reactions are summarized
in Table \ref{tab:6.Heburn}. As neutron capture cross sections increase with mass
number, neutron captures occur preferentially on the most abundant heavy
nucleus, i.e. $^{56}$Fe. Some intermediate mass nuclei, which are
produced in important quantities due to the $^{22}$Ne neutron source
and neutron capture chains based on it, are $^{21}$Ne, $^{22}$Ne,
$^{25}$Mg, $^{26}$Mg, $^{36}$S, $^{37}$Cl, $^{40}$K, and $^{40}$Ar.
The status of uncertainties in He-burning reactions, 
especially with respect to
$(\alpha ,n)$- and $(n,\g)$-reactions, is discussed in 
\cite{Kaeppeler.Gallino.ea:2011,Fabbraro.deBoer.ea:2020}, but see also the footnotes at the end of the subsection on H-burning.

\begin{table}[h!]
\centering
\caption{Major Reactions in Helium Burning}
\begin{tabular}{l}
\hline
\hline
(a) basic energy generation \\
\ $^{4}$He(2$\a,\g)^{12}$C$~~~^{12}$C($\a,\g)^{16}$O[($\a,\g)^{20}$Ne] \\
\noalign{\vskip2pt}
(b) neutron sources  \\
\noalign{\vskip2pt}
\ $^{14}$N($\a,\g)^{18}$F($e^+\nu)^{18}$O($\a,\g)^{22}$Ne
 $^{22}$Ne($\a$,n)$^{25}$Mg \\
\ $^{12}$C(p,$\g)^{13}$N($e^+\nu)^{13}$C($\a$,n)$^{16}$O \\
\noalign{\vskip2pt}
(c) high temperature burning with neutron sources  \\
\noalign{\vskip2pt}
\ $^{22}$Ne(n,$\g)^{23}$Ne($e^-\bar\nu)^{23}$Na(n,$\g)^{24}$Na
($e^-\bar\nu)^{24}$Mg \\
\ $^{20}$Ne(n,$\g)^{21}$Ne($\a$,n)$^{24}$Mg \\
\ further s-processing via neutron captures and $\b$-decays$~~~~~~
^{24}$Mg(n,$\g)^{25}$Mg etc. \\
\ production of heavy elements
$~~~^{56}$Fe(n,$\g)^{57}$Fe(n,$\g)^{58}$Fe etc.\\
\noalign{\vskip1pt} 
\hline
\hline
\end{tabular}
\label{tab:6.Heburn}
\end{table}

\subsection {Carbon, Neon, and Oxygen Burning}

\subsubsection{Carbon Burning}
\label{6.3.1}

We want to point out here that stars less massive
than about 8M$_\odot$ end their core evolution after helium burning.
They become (electron-)degenerate and do not contract further due to the
electron degeneracy pressure. Cores of more massive stars do contract, heat up
and induce the next burning stage.
Among the ashes of helium burning the nucleus with the smallest charge is 
$^{12}$C.  At densities of $\rho\approx10^5\gcm$ and
temperatures of $T\approx(6-8)\times 10^8$ K the Maxwell-Boltzmann velocity
distribution reaches energies sufficient to penetrate the Coulomb barrier
in the $^{12}$C+$^{12}$C reaction, which has two open particle channels
$$^{12}\ce\left(^{12}\ce,{\a\atop p}\right){^{20}{\rm Ne}\atop ^{23}{\rm
Na}}~~~~{ 4.62\mev\atop 2.24\mev.}$$ 
The products can interact with the released particles (protons and alphas)
$$^{23}{\rm Na}(p,\a)^{20}{\rm Ne}~~~~^{23}{\rm Na}(p,\g)^{24}{\rm Mg}~~~~
^{12}\ce(\a,\g)^{16}\oh.$$
Besides these reactions, which dominate the energy generation, we also list in
Table \ref{tab:6.Cburn} the reactions whose fluxes are down by up to a factor of 100.

\begin{table}[h!]
\centering
\caption{Major Reactions in Carbon Burning}
\begin{tabular}{l}
\hline
\hline
(a) basic energy generation   \\
\ $^{12}$C($^{12}$C,$\a)^{20}$Ne$~~~~^{12}$C($^{12}$C,p)$^{23}$Na \\
\ $^{23}$Na(p,$\a)^{20}$Ne$~~~^{23}$Na(p,$\g)^{24}$Mg$~~~^{12}$C($\a,\g)^{16}$O
\\
(b) fluxes $>10^2\times$(a) \\
\ $^{20}$Ne($\a,\g)^{24}$Mg$~~~~^{23}$Na($\a$,p)$^{26}$Mg(p,$\g)^{27}$Al \\
\ $^{20}$Ne(n,$\g)^{21}$Ne(p,$\g)^{22}$Na
($e^+\nu)^{22}$Ne($\a$,n)$^{25}$Mg(n,$\g)^{26}$Mg \\
\ $^{21}$Ne($\a$,n)$^{24}$Mg$~~~^{22}$Ne(p,$\g)^{23}$Na$~~~ 
^{25}$Mg(p,$\g)^{26}$Al($e^+\nu)^{26}$Mg \\
(c) low temperature, high density burning    \\
\ $^{12}$C(p,$\g)^{13}$N($e^+\nu)^{13}$C($\a$,n)$^{16}$O($\a,\g)^{20}$Ne \\
\ $^{24}$Mg(p,$\g)^{25}$Al($e^+\nu)^{25}$Mg \\
\ $^{21}$Ne(n,$\g)^{22}$Ne(n,$\g)^{23}$Ne($e^-\bar\nu)^{23}$Na(n,$\g)^{24}
$Na($e^-\nu)^{24}$Mg + s-processing\\
\hline
\hline
\end{tabular}
\label{tab:6.Cburn}
\end{table}
The Q-value of the proton capture reaction on $^{12}$C is 1.93MeV.
Eq.(\ref{eq:reac-photo}), when utilizing the detailed balance relation between forward capture and reverse photodisintegration, results in the fact that the photodisintegration reactions are faster than 
the inverse capture reactions if $kT>1/24 Q$. This means that in high 
temperature carbon burning, close to $8\times 10^8$K, the photodisintegration
of $^{13}$N will win over the $\b^+$-decay and the reaction sequence (c) in
Table \ref{tab:6.Cburn} will not be of any importance. However, in low temperature carbon
burning the photodisintegration is slow, and we have the $^{13}$C neutron
source similar to hot helium burning, with the protons being supplied from the
carbon fusion reaction.
Only in this latter case an increase in the total neutron/proton ratio
occurs, which can decrease
$Y_e$ down to 0.4975. For burning conditions with higher temperatures
$Y_e$ from helium burning remains
unchanged.

Besides the uncertainties in the  measurements of reactions overlapping with
Table \ref{tab:6.Heburn}, one has to discuss the behavior of the dominant fusion
reaction $^{12}$C+$^{12}$C \citep[see e.g. for early and recent investigations][]{rolfs.rodney:1988,Caughlan.Fowler:1988,Tang.Ru:2022,Heine.ea:2022,Monpribat.ea:2022}.

\subsubsection{Neon Burning}
\label{sec:6.3.2}

The most abundant nuclei after carbon burning are $^{16}$O, $^{20}$Ne, and
$^{24}$Mg. The reaction
$^{16}$O($\a,\g)^{20}$Ne has a Q-value of only 4.73MeV. The typical reaction
Q-value for reactions among stable nuclei is about 8-12MeV. This means, that in the temperature range
of 1-2MeV (following the relation $kT>1/24 Q$),
alpha particles will be liberated by the photodisintegration of $^{20}$Ne,
and can be used to rearrange nuclei
$$^{20}{\rm Ne}(\g,\a)^{16}\oh$$
$$^{20}{\rm Ne}(\a,\g)^{24}{\rm Mg}(\a,\g)^{28}{\rm Si}.$$

The typical conditions for neon burning in stars are $T\approx (1.2-1.4)
\times 10^9$K and $\rho\approx 10^6\gcm$. Stars in the mass range
8-10M$_\odot$
appear to not reach this stage to burn Ne in a stable fashion, but their cores become degenerate and they
contract to high densities at low temperatures.
In that case the
electron Fermi energies start to become important and electron captures
start to occur. The electrons turn degenerate with a Fermi energy for the completely degenerate non-relativistic case and
$n_e=(\rho N_AY_e)^{2/3}$ 
\begin{align}
E_F&={\hbar^2\over 2m_e}(3\pi^2)^{2/3}ne^{2/3} \nonumber \\
E_F(\rho Y_e&=10^7\gcm)=0.75\mev \label{eq:6.13} \\
E_F(\rho Y_e&=10^9\gcm)=4.7\mev.\nonumber
\end{align}
That means that electron captures, which are energetically prohibited
$e^-+(Z,A)\rightarrow(Z-1,A)+\nu$ with a negative nagative Q-value,
can become possible and lead to an enhanced ``neutronization'' of the 
astrophysical plasma, in addition to the role of beta decays and electron
captures with positive Q-values.
In degenerate Ne-O-Mg cores electron captures on $^{20}$Ne and $^{24}$Mg
cause the loss of degeneracy pressure support and introduce a collapse 
rather than only a contraction, which shortens all further burning stages
on a collapse time scale \citep[for detailed discussions see][]{Kirsebom.Jones.ea:2019,Leung.Nomoto.ea:2020}. 

More massive stars pass through the main neon burning reactions, as discussed
above. In addition, many other reactions can occur with fluxes much smaller 
than for those reactions
that dominate the energy generation. We present a list in Table \ref{tab:6.Neburn}.
However, there is not any important change in $Y_e$ for these
massive stars.
For an overview discussing experimental and theoretical evaluation of alpha-induced reactions from Ne and Mg to Ca see \cite{Rauscher.Thielemann.Goerres.Wiescher:2000}.

\begin{table}[h!]
\centering
\caption{Major Reactions in Neon Burning}
\begin{tabular}{l}
\hline
\hline
(a) basic energy generation   \\
\ $^{20}$Ne($\g,\a)^{16}$O$~~~~^{20}$Ne($\a,\g)^{24}$Mg($\a,\g)^{28}$Si\\
(b) fluxes $>10^2\times$(a) \\
\ $^{23}$Na(p,$\a)^{20}$Ne$~~~~^{23}$Na($\a$,p)$^{26}$Mg($\a$,n)$^{29}$Si \\
\ $^{20}$Ne(n,$\g)^{21}$Ne($\a$,n)$^{24}$Mg(n,$\g)^{25}$Mg($\a$,n)$^{28}$Si
\\
\ $^{28}$Si(n,$\g)^{29}$Si(n,$\g)^{30}$Si \\
\ $^{24}$Mg($\a$,p)$^{27}$Al($\a$,p)$^{30}$Si \\
\ $^{26}$Mg(p,$\g)^{27}$Al(n,$\g)^{28}$Al($e^-\bar\nu)^{28}$Si\\
(c) low temperature, high density burning    \\
\
$^{22}$Ne($\a$,n)$^{25}$Mg(n,$\g)^{26}$Mg(n,$\g)^{27}$Mg($e^-\bar\nu)^{27}$Al
\\
\ $^{22}$Ne left from prior neutron-rich carbon burning  \\
\hline
\hline
\end{tabular}
\label{tab:6.Neburn}
\end{table}

\subsubsection{Oxygen Burning}
\label{sec:6.3.3}

The oxygen, produced in helium burning
$[^4\he(2\a,\g)^{12}\ce(\a,\g)^{16}\oh]$,
in carbon burning $[^{12}\ce(\a,\g)^{16}\oh$ with $\a's$ from
$^{12}\ce(^{12}\ce,\a)^{20}$Ne],
and in neon burning $[^{20}$Ne($\g,\a)^{16}\oh]$, is still unburned and the
nucleus with the smallest charge. At $\rho\approx 10^7\gcm$ and
$T\approx(1.5-2.2)\times 10^9$K the Coulomb barrier for oxygen fusion can
be overcome, leading to the major reactions

\begin{align}
^{16}\oh(^{16}\oh,p)^{31}\pe\qquad&7.676\mev\nonumber\\
^{16}\oh(^{16}\oh,\a)^{28}\si\qquad&9.593\mev\\
^{16}\oh(^{16}\oh,n)^{31}\es(\b^+)^{31}\pe\qquad&1.459\mev\nonumber\\
^{31}\pe(p,\a)^{28}\si~~{\rm etc.}\ldots\nonumber
\end{align}

In high density oxygen burning ($\rho>2\times 10^7\gcm$) two electron
capture reactions become important and lead to a decrease in $Y_e$
$$^{33}\es(e^-,\nu)^{33}\pe~~~~^{35}{\rm Cl}(e^-,\nu)^{35}\pe.$$
$Y_e$ can decrease to 0.495-0.4825.
We list in Table \ref{tab:6.Oburn} the major and minor reaction sequences.
Many individual reactions of intermediate mass nuclei in oxygen and
silicon burning have been analyzed in the past. Such experiments
usually find quite good agreement with statistical model
(Hauser-Feshbach) calculations, which will be discussed in the next section. Here we still want to point to the tremendous experimental efforts undertaken for determining solid cross section data.

\begin{table}[h!]
\centering
\caption{Major Reactions in Oxygen Burning}
\begin{tabular}{l}
\hline
\hline
(a) basic energy generation   \\
\ $^{16}$O($^{16}$O,$\a)^{28}$Si$~~~~^{16}$O($^{12}$O,p)$^{31}$P$~~~~
^{16}$O($^{16}$O,n)$^{31}$S($e^+\nu)^{31}$P \\
\ $^{31}$P(p,$\a)^{28}$Si($\a,\g)^{32}$S\\
\ $^{28}$Si($\g,\a)^{24}$Mg($\a$,p)$^{27}$Al($\a$,p)$^{30}$Si \\
\ $^{32}$S(n,$\g)^{33}$S(n,$\a)^{30}$Si($\a,\g)^{34}$S \\
\ $^{28}$Si(n,$\g)^{29}$Si($\a$,n)$^{32}$S($\a$,p)$^{35}$Cl \\
\ $^{29}$Si(p,$\g)^{30}$P($e^+\nu)^{30}$Si \\
electron captures \\
$^{33}$S($e^-,\nu)^{33}$P(p,n)$^{33}$S \\
$^{35}$Cl($e^-,\nu)^{35}$S(p,n)$^{35}$Cl \\
(b) high temperature burning \\
\ $^{32}$S($\a,\g)^{36}$Ar($\a$,p)$^{39}$K \\
\ $^{36}$Ar(n,$\g)^{37}$Ar($e^+\nu)^{37}$Cl\\
\ $^{35}$Cl($\g$,p)$^{34}$S($\a,\g)^{38}$Ar(p,$\g)^{39}$K(p,$\g)^{40}$Ca \\
\ $^{35}$Cl($e^-,\nu)^{35}$S($\g$,p)$^{34}$S \\
\ $^{38}$Ar($\a,\g)^{42}$Ca($\a,\g)^{46}$Ti \\
\ $^{42}$Ca($\a$,p)$^{45}$Sc(p,$\g)^{46}$Ti \\
(c) low temperature, high density burning    \\
\ $^{31}$P($e^-\nu)^{31}$S$~~~~^{31}$P(n,$\g)^{32}$P \\
\ $^{32}$S($e^-,\nu)^{32}$P(p,n)$^{32}$S \\
\ $^{33}$P(p,$\a)^{30}$Si \\
\hline
\hline
\end{tabular}
\label{tab:6.Oburn}
\end{table}

\medskip
\subsection{ Silicon Burning}
\label{sec:6.4}
When temperatures reach $(3-4)\times 10^9$ K, every nucleus is connected to
each other by reaction links which are open in both directions: (i) capture 
reactions due to high enough temperatures to overcome Coulomb barriers, (ii)
photodisintegrations due to high enough temperatures and high energy photons
in the Planck distribution. This situation,  initiated by 
photodisintegration reactions on dominant $^{28}$Si, which produce
protons, alpha-particles and neutrons, finally leads to an equilibrium 
where the abundances depend only on the nuclear mass (binding energy), density 
and temperature. Then the composition can be described by nuclear
statistical equilibrium (NSE) and expressed by Eq.(\ref{eq:6.14}), which results from utilizing the chain of abundance ratios as given in Eq.(\ref{eq:2}) (an alternative derivation results from utilizing the chemical potentials of neutrons, protons and nucleus $(Z,N)$ from a Maxwell-Boltzmann distribution and solving for the equation $Z\mu_p+N\mu_n=\mu_{Z,N}$).

\begin{equation}
Y_{Z,N}=G_{Z,N}(\rho N_A)^{A-1}{A^{3/2}\over 2^A}\left({2\pi\hbar^2
\over m_ukT}\right)^{3/2(A-1)}\exp(B_{Z,N}/kT)Y^N_nY^Z_p.
\label{eq:6.14}
\end{equation}

Electron captures, which occur on longer time scales than particle captures
and photodisintegrations, are not in equilibrium and have to be followed
explicitly. Thus, the NSE has
to be solved for a given $\rho(t)$, $T(t)$, and $Y_e(t)$. This leads to two
equations and a composition, resulting from Eq.(\ref{eq:6.14}) and 
a given $Y_e$, determined by

\begin{align}
\sum_i A_i Y_i&=Y_n+Y_p+\sum_{Z,N}
(Z+N)Y_{Z,N}(\rho,T,Y_n,Y_p)=1\nonumber\\
\sum_i Z_i Y_i&=Y_p+\sum_{Z,N} Z~Y_{Z,N}(\rho,T,Y_n,Y_p)=Y_e.\label{eq:6.16}
\end{align}

In general, very high densities favor large nuclei, due to the high power of
$\rho^{A-1}$, while very high temperatures favor light nuclei, due to 
$(kT)^{-3/2(A-1)}$. In the intermediate regime, occurring in hydrostatic
stellar
evolution, $\exp (E_B/kT)$ favors tightly bound nuclei with the highest
binding
energies in the mass range $A=50-60$ (of the Fe-group), but depending 
upon the
given $Y_e$. The width of the composition distribution is determined by the
temperature.
$Y_e$ is determined by $\b^+$-decays and electron captures in Si-burning.
Close to the end of core silicon burning, we have $Y_e\approx 0.46$.

\begin{figure}[h!]
    \centering
     \includegraphics[width=0.6\linewidth]{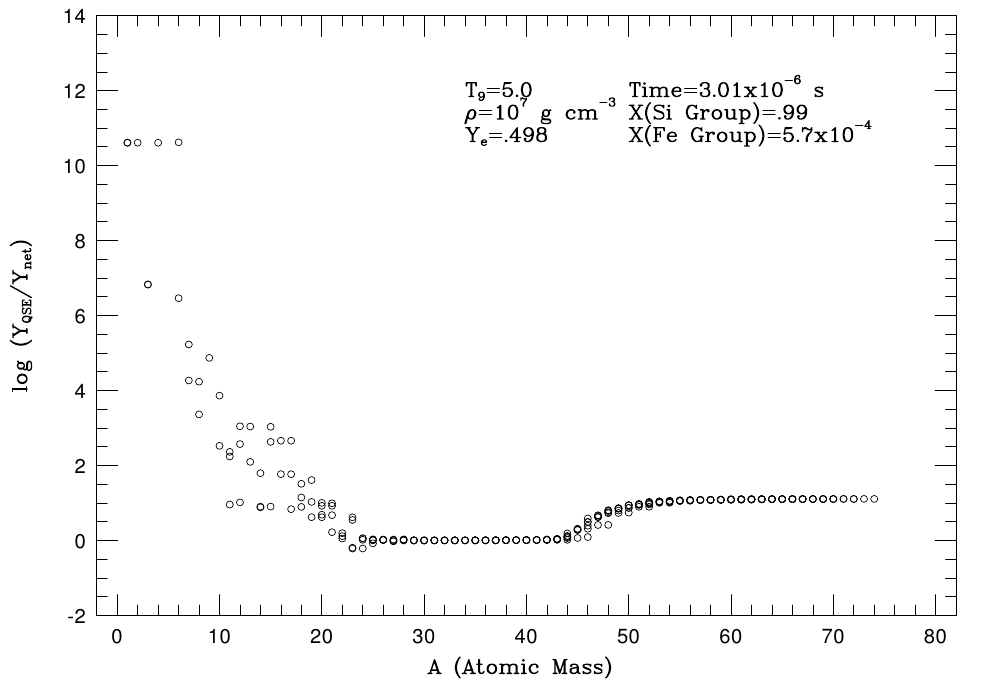}
    \includegraphics[width=0.6\linewidth]{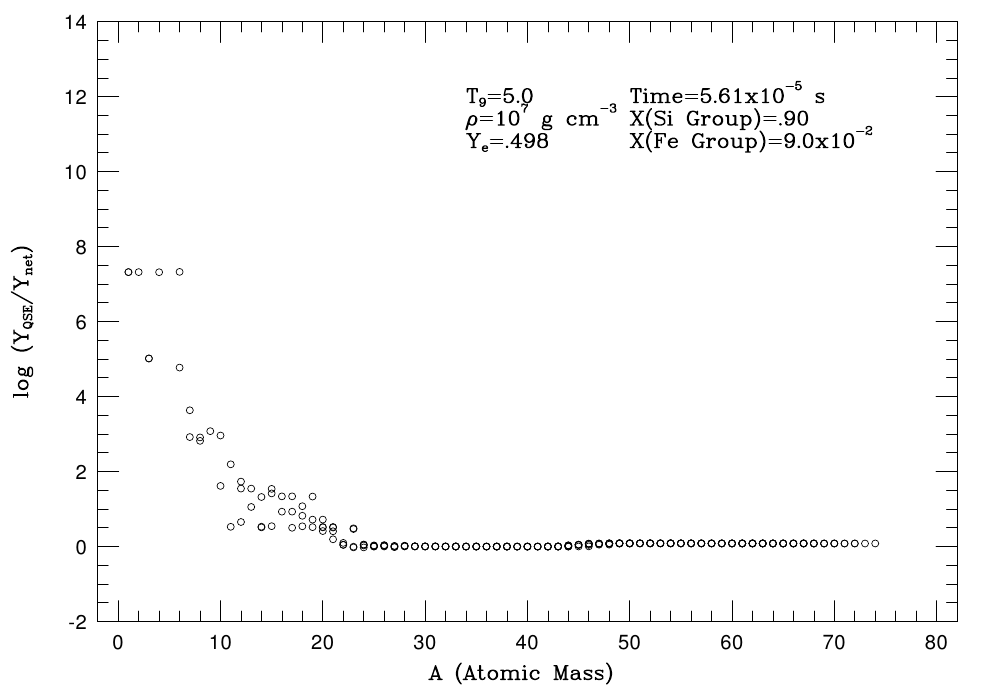}
    \caption{Shown is the logarithmic ratio of equilibrium abundance predictions in comparison to results from full network calculations at two points in time when the mass fraction in the Si-group is still 99 and later 90\%. Initially only the Si QSE-group is in full equilibrium (=0), while the light nuclei created in photodisintegration reactions and the heavier Fe-group nuclei created by capture reactions are still underabundant in comparison to equilibrium. However, we see constant lines for other QSE groups, i.e. the light group of n, p, $^{4,6}$He and the Fe-group. The Fe group approaches equilibrium fast already when $X(Si)$=0.9 while it will take a longer timescale by another order of magnitude until also the lighter nuclei have reached
    equilibrium. Image reproduced with permission from \cite{Hix.Thielemann:1996}, copyright by AAS. }
    \label{fig:nseapproach}
\end{figure}

However, this final stage of Si-burning, resulting in a complete chemical equilibrium or NSE, is initially hampered by slow reactions, especially due to small Q-values of reactions into closed neutron or protons shells. This leads initially to quasi-equilibrium groups (QSE), e.g. around $^{28}$Si and a group centering on Fe or Ni nuclei, which are both within their groups in a complete equilibrium with abundance ratios as in Eq.(\ref{eq:2}), but the connections across the closed shells $N=Z=20$ are hindered. Then the relative ratios of neighboring nuclei in each of the groups are in equilibrium, but the sum of abundances in each of the two groups is not identical to an NSE distribution. It is only approached by starting out with the QSE around Si dominating and, as a function of time, both - the Si and the Fe QSE groups - attain their final NSE values.
\begin{SCfigure}
     \includegraphics[width=0.6\linewidth]{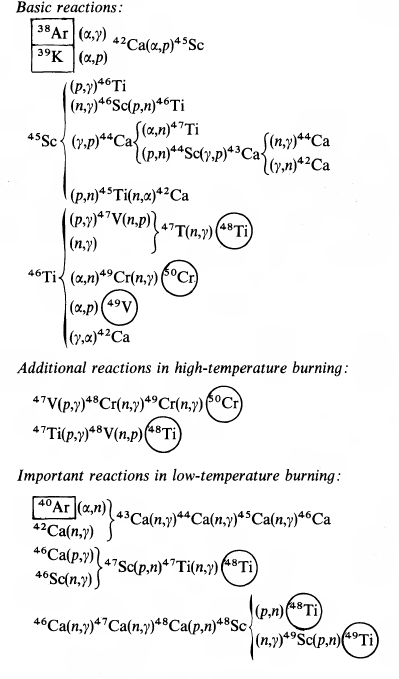}
\caption{The major reactions which connect the Si QSE-group and the Fe QSE-group during hydrostatic stellar Si-burning for $Y_e$=0.498 from an early analysis \cite{Thielemann.Arnett:1985}. Dependent on the stellar mass this takes place for higher central temperatures (high mass stars) or slightly lower temperatures (massive stars in the lower mass interval $>$10M$_\odot$ which still experience central Si-burning). Members of the Si-group are indicated by a box, members of the Fe-group by a circle. Image reproduced with permission from \cite{Thielemann.Arnett:1985}, copyright by AAS. }
    \label{fig:qseconnection}
\end{SCfigure}

This behavior has been discussed in \cite{Hix.Thielemann:1996}. Here we present a few illustrations clarifying this background. While temperatures of $T_9$=5 are in stellar evolution only reached at the end of Si-burning, we give here an example of a plasma consisting initially only of Si isotopes with a composition representing $Y_e$=0.498. We show the evolution at two points in time around $3\times 10^{-6}$ and $5.6\times 10^{-5}$s. Light particles are created by photodisintegration reactions on Si and heavier nuclei are produced by captures of these light particles. This means that initially these light particles and the heavier Fe-group are underabundant in comparison to a full NSE. This can be seen in the two panels of Fig.\ref{fig:nseapproach}. 

\begin{figure}[h!]
     \includegraphics[width=0.5\linewidth]{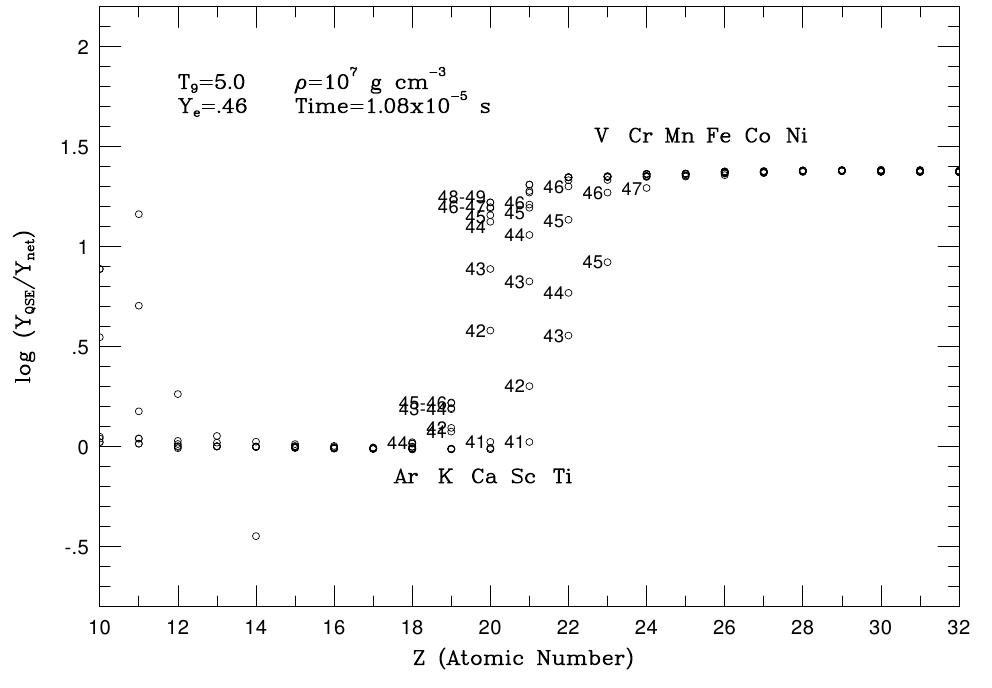}
     \includegraphics[width=0.5\linewidth]{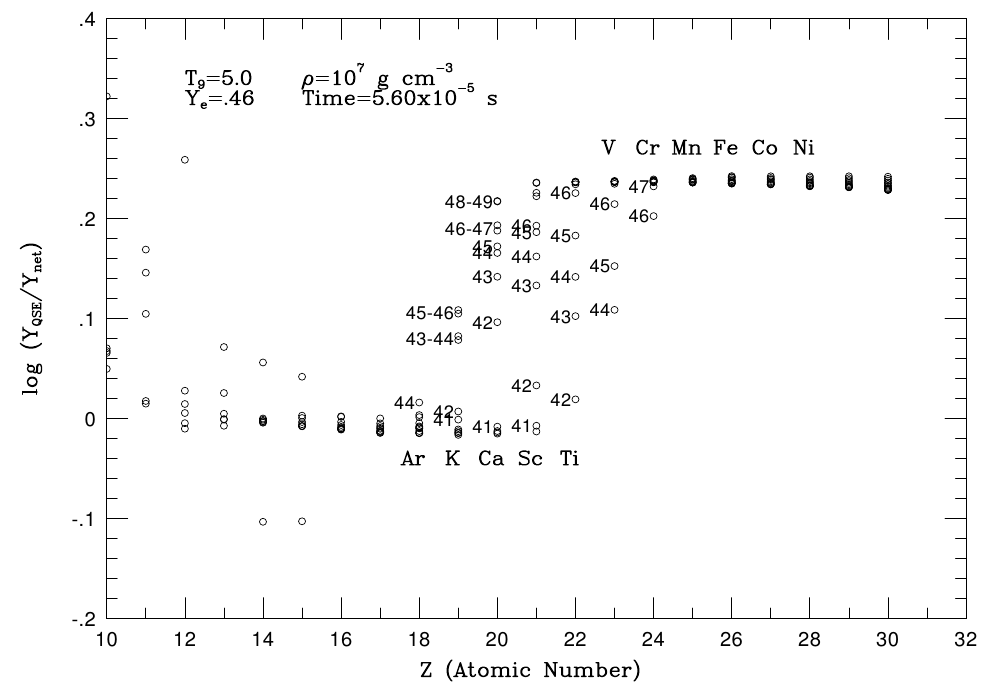}
\caption{The deviation from equilibrium decreases as a function of time, being already very close to equilibrium for $t\approx5\times 10^{-5}$s. But it can be realized as well, that there exist still a (small) deviation between the two groups and that the nuclei out of equilibrium between the two groups vary as a function of time, increasing with $Z$ (and being relatively neutron-rich for $Y_e$=0.46). In both cases nuclei close to the borders of the shell closures $N=Z=20$ are the ones which still hinder to reach an equilibrium fast due to their slow reaction rates for reactions with small Q-values.Image reproduced with permission from \cite{Hix.Thielemann:1996}, copyright by AAS. }
    \label{fig:qseconnection2}
\end{figure}
\cite{Thielemann.Arnett:1985} made an early attempt for such $Y_e$-conditions to identify the important (slow) reactions which connect 
the two groups (see Fig.\ref{fig:qseconnection}). 
In a more extended study \cite{Hix.Thielemann:1996} analyzed the nuclei connecting the two QSE-groups. We can see the nuclei positioned between the two QSE-groups for two sets of conditions with $Y_e$=0.46, but at two points in time, close to $t\approx1$ and $5\times 10^{-5}$s. We notice that the deviation from equilibrium decreases as a function of time, being already very close to equilibrium in the latter case, but it can be realized as well, that there exist still a (small) deviation between the two groups and that the nuclei out of equilibrium between the two groups vary as a function of time, with increasing $Z$ and are relatively neutron-rich for $Y_e$, see Fig.\ref{fig:qseconnection2}.

\begin{figure}[h!]
     \includegraphics[width=0.49\linewidth]{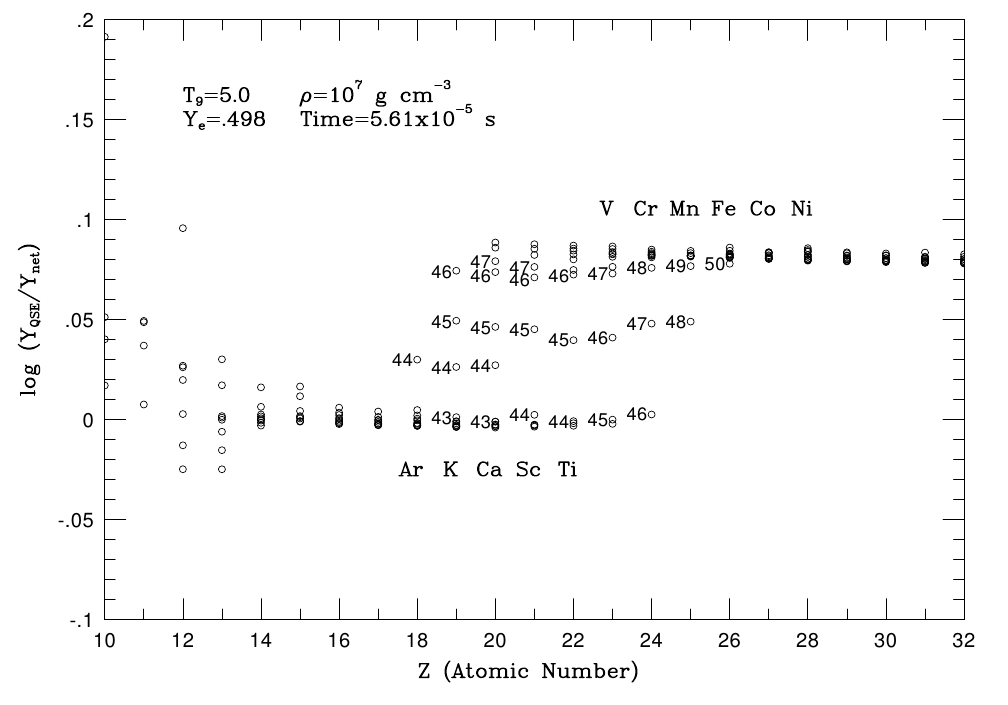}
     \includegraphics[width=0.51\linewidth]{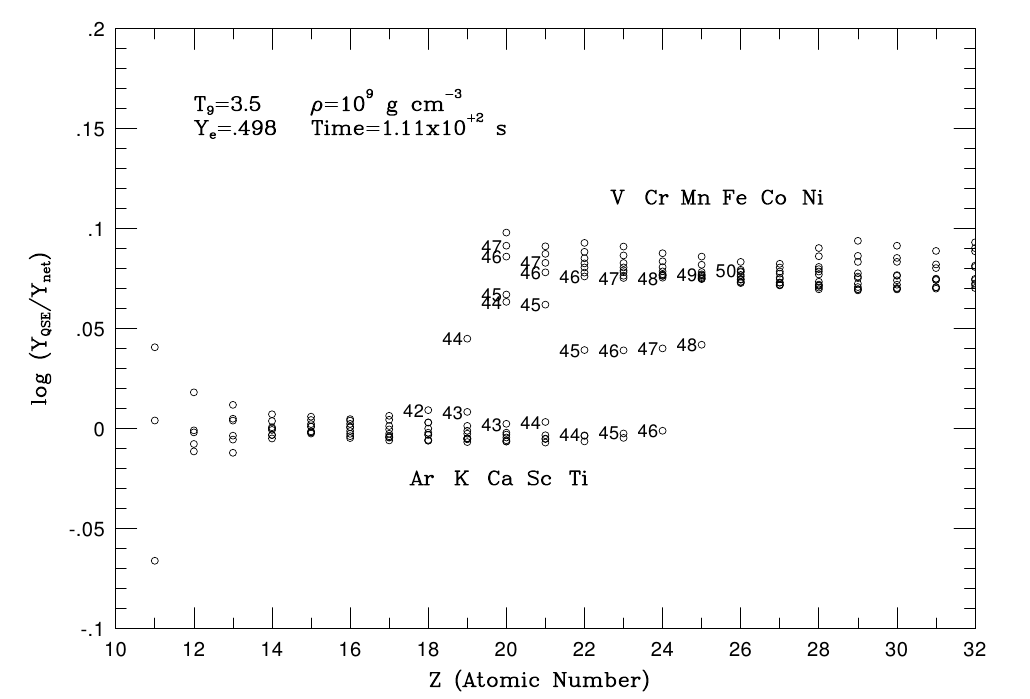}
\caption{Two results for $Y_e$=0.498 but at quite different temperatures and points in time. Similar deviations from equilibrium prevail for $t\approx5\times 10^{-6}$s and in contrast for 111s. In both cases finally an NSE will be reached, but either for conditions close to explosive Si-burning or otherwise for conditions which are similar to the initial ignition conditions of hydrostatic Si-burning in massive stars. Image reproduced with permission from \cite{Hix.Thielemann:1996}, copyright by AAS.}
    \label{fig:qseconnection3}
\end{figure}

Especially this last subsection, Si-burning, but also the previously discussed O-burning contain an overwhelming number of reactions of intermediate mass and heavy nuclei which need to be known and cannot all be provided by experiments. Thus, a reliable method for theoretical predictions is needed. The following section (after also going through explosive burning stages in the next subsection) will focus on the statistical model or Hauser-Feshbach approach, that can be applied if a sufficiently large density of excited state levels in the compound nucleus exists.

\section{Nuclear burning in explosive environments}
\medskip

Many of the hydrostatic burning processes discussed in earlier subsections can
occur also under explosive conditions at much higher temperatures and on
shorter timescales. The major reactions remain still the same in many
cases, but often the beta-decay half-lives of unstable products are
longer than the timescales of the explosive processes under investigation.
This requires in general the additional knowledge of nuclear cross
sections for unstable nuclei. (For the further influence of weak interactions
in hydrostatic and explosive burning see \cite{Langanke.Martinez:2021} and the chapter by Famiano et al.).

Extensive calculations of explosive carbon, neon, oxygen, and silicon
burning, appropriate for supernova explosions, have already been performed 
in the late 60s and early 70s with the accuracies possible in those days,
before detailed stellar modeling became available 
\citep[for present-day results see e.g.][and the chapters by Seitenzahl \& Pakmor, C. Fr\"ohlich, S. Wanajo, 
M. Obergaulinger, and A. Bauswein \& H.-T. Janka]{Curtis.ea:2019}.
Besides minor additions of $^{22}$Ne after He-burning
(or nuclei which originate from it in later burning stages), the fuels
for explosive nucleosynthesis consist mainly of alpha-particle nuclei like
$^{12}$C, $^{16}$O, $^{20}$Ne, $^{24}$Mg, or $^{28}$Si. Because the
timescale of explosive processing is very short (a fraction of a second
to several seconds), only few beta-decays can occur during 
explosive nucleosynthesis events (unless highly unstable species are produced), resulting in heavier nuclei, again with 
N$\approx$Z. However, a spread of nuclei around a line of N$=$Z
is involved and many reaction rates for unstable nuclei have to be known. 
Dependent on the temperature, explosive burning produces intermediate to 
heavy nuclei.

Two processes encounter nuclei far from stability where either a large
supply of neutrons or protons is available. In those cases, the
r-process and the rp-process (i.e. explosive hydrogen burning), nuclei
close to the neutron and proton drip lines can be produced and beta-decay
timescales can be short in comparison to the process timescales.
We will not discuss these nucleosynthesis processes in detail here, but refer to the 
chapters by M. Wiescher and M. Pf\"utzner \& C. Mazzocchi for reactions involving
proton-rich nuclei and the chapters by S. Goriely and S. Nishimura for reactions
involving neutron-rich nuclei in r-process environments.
Independent of these individual aspects, for all the intermediate and heavy nuclei
we will present (after a short discussion of these explosive nucleosynthesis environment) theoretical methods to determine
nuclear reaction cross sections and reaction rates.

While most of the explosive burning processes are to some extent (with the exceptions mentioned above and below)
"cousins" of their hydrostatic versions, involving just much shorter timescales and
a more extended set of nuclei, this is different for Si-burning which we will discuss 
in a separate subsection. A possible connection between explosive Si-burning and the beginning of an 
r-process is pointed out as well.

\subsection{Burning timescales for explosive He, C, Ne, O, and Si-burning}

In stellar evolution, burning timescales are dictated by the energy loss
timescales of stellar environments. Processes like hydrogen and helium
burning, where the stellar energy loss is dominated by the photon
luminosity, choose temperatures with energy generation rates 
equal to the radiation losses. For the later burning stages neutrino losses 
play the dominant role among cooling processes and the burning timescales 
are determined by temperatures where neutrino losses are equal to the
energy generation rate. These criteria led to the narrow bands in
temperatures and densities discussed in previous subsections on hydrostatic burnin stages in stellar evolution.

Explosive events are determined by hydrodynamic equations which provide
different temperatures or timescales for the burning of available fuel.
We can generalize the question by defining a burning timescale, dependent on 
whether the key reaction is a fusion reaction or a photodisintegration, responsible
for the destruction of the major fuel nuclei $i$
\begin{align}
\tau_i=|{Y_i \over \dot Y_i}|&= {1\over {\rho N_A < \sigma v >(T) Y_{fuel}}} \label{eq:desttau}\\
{\rm or}&={1 \over \lambda_\gamma (T)} \nonumber
\end{align}

\begin{figure}
    \centering
    \includegraphics[width=0.6\linewidth,angle=-90]{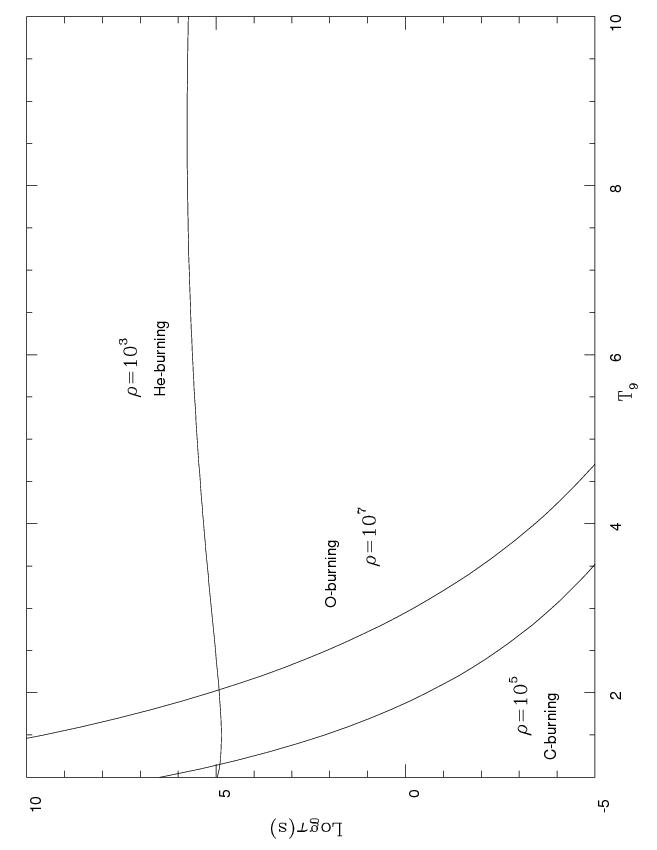}
    \includegraphics[width=0.6\linewidth,angle=-90]{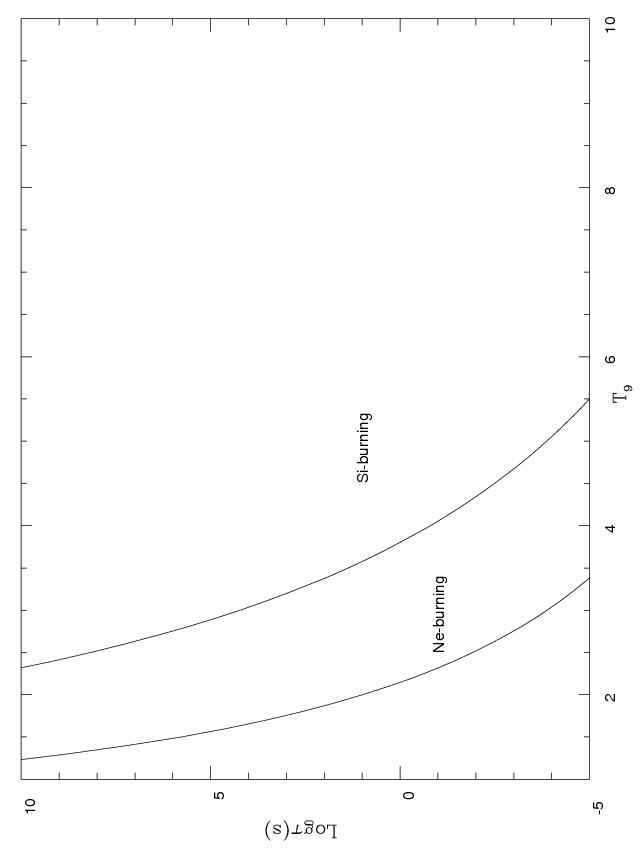}
\caption{Burning time scales for fuel destruction of He-, C-, Ne-, O-,
and Si-burning as a function of temperature, as defined in Eq.(\ref{eq:desttau}
and generalized to photodisintegrations and
three-body reactions, accordingly. A 100 fuel mass fraction
was assumed. When rewriting
Eq.(\ref{eq:desttau}) for $^4$He, the destruction of three identical particles
has to be considered.
The density-dependent burning time scales are labeled with the
chosen typical density. They scale linearly for C- and O-burning
and quadratically for He-burning. Notice that the almost constant
He-burning time scale beyond $T_9$=1 has the effect that efficient
destruction on explosive time scales can only be attained for high
densities. Image reproduced with permission from \cite{thielemann18}, copyright by Springer Nature.}
\label{fig:exploburn}
\end{figure}
These timescales for the fuels $i\in$ H, He, C, and O are determined
by the major fusion destruction reaction. They are in all cases temperature
dependent and for a fusion process they are also density
dependent. Ne- and Si-burning, which are dominated by $(\gamma,\alpha)$
destruction of $^{20}$Ne and $^{28}$Si, have timescales only determined by
the burning temperatures. The temperature dependencies are typically
exponential, due to the functional form of the corresponding $N_A<\sigma
v>$. The density dependencies are linear for fusion reactions, as seen from
Eq.(\ref{eq:networkequ}), with the exception of He-burning.
We have plotted these burning timescales as a function of temperature
(see Fig.\ref{fig:exploburn}), assuming a fuel mass
fraction of 1. The curves for (also) density dependent burning processes
are labeled with a typical density.
He-burning has a quadratic density dependence, C- and O-burning
depend linearly on density.
If we take typical explosive burning timescales to be of the order of
seconds (e.g. in supernovae), we see that one requires temperatures to
burn essential parts of the fuel in excess of 4$\times 10^9$K (Si-burning),
3.3$\times 10^9$K (O-burning), 2.1$\times 10^9$K (Ne-burning), and
1.9$\times 10^9$K (C-burning). Beyond $10^9$K He-burning is
determined by an almost constant burning time scale. We see that
essential destruction on a time scale of 1s is only possible for
densities $\rho$$>$$10^5$g cm$^{-3}$. This is usually not encountered
in He-shells of massive stars.

\subsection{Special features of explosive Si-burning}

\begin{figure}
    \centering
    \includegraphics[angle=-90,width=0.8\linewidth]{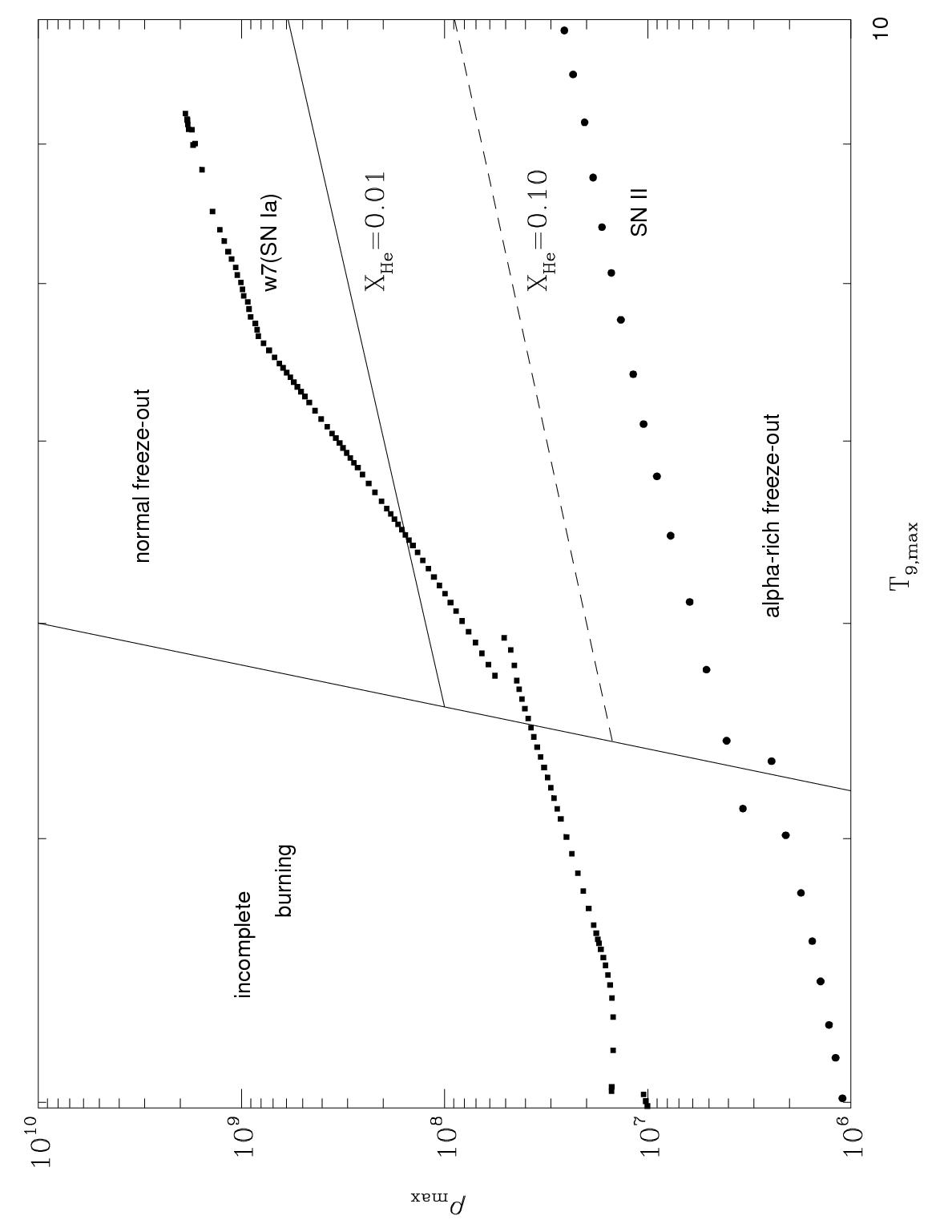}
    \caption{Division of the $\rho_p - T_p$-plane into
incomplete and complete Si-burning with normal and alpha-rich freeze-out.
Contour lines of constant $^4$He mass fractions in complete burning are
given for levesl of 1 and 10\%. They coincide with lines of constant radiation
entropy per gram of matter. For comparison also the maximum
$\rho-T$-conditions for individual mass zones of type Ia and type II (core-collapse) supernovae
are indicated. Image reproduced with permission from \cite{thielemann18}, copyright by Springer Nature. }
    \label{fig:explosi}
\end{figure}

Zones which experience temperatures in excess of 4.0--5.0$\times 10^9$K
undergo explosive Si-burning. For $T$$>$5$\times 10^9$K essentially all 
Coulomb barriers can be overcome and a nuclear statistical equilibrium is
established. Such temperatures lead to complete Si-exhaustion and produce 
Fe-group nuclei. The doubly-magic nucleus $^{56}$Ni, with the largest
binding energy per nucleon for N=Z, is formed with a dominant abundance
in the Fe-group in case $Y_e$ is larger than 0.49.
Explosive Si-burning can be divided into three different regimes:
(i) incomplete Si-burning and complete Si-burning with
either a (ii) normal or an (iii) alpha-rich freeze-out.
Which of the three regimes is encountered depends on the peak temperatures
and densities attained during the passage of supernova shock front \citep[see Fig.20
in][and Fig.\ref{fig:explosi} for applications to supernova
calculations]{Woosley.Arnett.Clayton:1973}.
One recognizes that SNe Ia and SN II (core-collapse supernovae) experience different regions
of complete Si-burning. 

The most abundant nucleus in the normal and alpha-rich freeze-out is
$^{56}$Ni, in case the neutron excess is
small, i.e., $Y_e$ is larger than 0.49. It is evident that the electron
fraction $Y_e$ has a major imprint on the final composition.
Nuclei with a $Z/A$-value corresponding to the global $Y_e$ show the largest
abundances, combined with a maximum in binding energy. The even-Z-even-N nuclei 
show the largest abundances due to their large bindung energies.
In an alpha-rich freeze-out final alpha-captures (or rather the shift of the 
upper QSE-group to heavier nuclei, based on the large $^4$He abundance) play a dominant
role for the less abundant nuclei, tranforming e.g. $^{56}$Ni, $^{57}$Ni, 
and $^{58}$Ni into $^{60}$Zn, $^{61}$Zn, and $^{62}$Zn  and leaving
trace abundances of 
$^{32}$S, $^{36}$Ar, $^{40}$Ca, $^{44}$Ti, $^{48}$Cr, $^{52}$Fe,
$^{54}$Fe, and $^{55}$Co. In that case the major NSE nuclei
$^{56}$Ni, $^{57}$Ni, and $^{58}$Ni get depleted when the remaining
alpha fraction increases, while all other species mentioned above
increase.  At high temperatures in
complete Si-burning, before freeze-out or in a normal freeze-out, the
abundances are in a full NSE. 
An alpha-rich freeze-out
occurs generally at low densities when the triple-alpha reaction,
transforming $^4$He into $^{12}$C, is not fast enough to keep the
He-abundance in equilibrium during the fast expansion and cooling in 
explosive events (see Fig.\ref{fig:explosi}). This also leads to a slow supply of
carbon nuclei still during freeze-out, leaving traces of alpha
nuclei, which did not fully make their way up to $^{56}$Ni.

Incomplete Si-burning is characterized 
by peak temperatures of $4-5\times 10^9$K. Temperatures are not high enough
for an efficient bridging of the bottle neck above the proton magic number
Z=20 by nuclear reactions. Besides the dominant fuel
nuclei $^{28}$Si and $^{32}$S, we find the alpha-nuclei $^{36}$Ar and
$^{40}$Ca being most abundant. 
Partial leakage through the bottle neck above Z=20 produces
$^{56}$Ni and $^{54}$Fe as dominant abundances in the Fe-group.
Smaller amounts of $^{52}$Fe, $^{58}$Ni,
$^{55}$Co, and $^{57}$Ni are encountered. 
In a superposition of (a) incomplete Si-burning, (b) complete
Si-burning with alpha-rich freeze-out, and (c) explosive O-burning
supernova explosions can provide good fits to solar abundances.
For recent features of the results of explosive burning in core-collapse supernovae see e.g. \cite{Curtis.ea:2019,Ghosh.ea:2022} and in type Ia supernovae see e.g. \cite{Lach.ea:2020,Leung.Nomoto:2022}.
\medskip

\subsection{The r-process}  

The operation of an r-process is characterized by the fact that
10 to 100 neutrons per seed nucleus in the Fe-peak (or beyond in an alpa-rich freeze-out) have to
be available to form all heavier r-process nuclei by neutron capture. 
This translates into a $Y_e$=0.15-0.3. Such a 
high neutron excess can only by obtained through capture of energetic
electrons 
(on protons or nuclei) which have to overcome large negative Q-values.
This can be achieved by degenerate electrons with large Fermi
energies and requires a compression to densities of
$10^{11}-10^{12}$g cm$^{-3}$, with a beta equilibrium between electron 
captures and $\beta^-$-decays as found in supernova core collapse and
the forming neutron stars. 

Another option is an extremely alpha-rich freeze-out in complete
Si-burning with moderate and $Y_e$'s (even $>$0.42). After the freeze-out of charged particle reactions
in matter which expands from high temperatures but relatively low
densities, 70, 80, 90 or 95\% of all matter can be locked into $^4$He with
N=Z. Figure \ref{fig:explosi} showed the onset of such an extremely alpha-rich
freeze-out by indicating contour lines for He mass fractions
of 1 and 10\%. These contour lines correspond to $T_9^3/\rho$=const,
which is proportional to the entropy per gram of matter of a radiation
dominated gas. Thus,
the radiation entropy per gram of baryons can be
used as a measure of the remaining He mass-fraction. In such high
entropy conditions all excess neutrons are left with the remaining mass 
fraction of Fe-group (or heavier) nuclei and neutron captures can proceed 
to form the heaviest r-process nuclei. A recent general review of the functioning of an r-process, the nuclear physics input, astrophysical sites and related observations can be found in \cite{Cowan.Sneden.ea:2021}.

\subsection{Explosive H-burning and proton-rich nuclei}

We will not discuss here in detail explosive hydrogen burning which takes place in novae and X-ray bursts, as their nucleosynthesis contribution to our known abundance pattern is not significant, however contributions to $^7$Li, $^{15}$N, and $^{17}$O, $^{22}$Na, $^{26}$Al are reported \citep[see e.g.][]{Cescutti.Molaro:2019,Vasini.ea:2022,Vasini.ea:2023}. Novae, due to high temperatures attained in H-burning, burn H in the accreted material not via the well-known CNO-cycle, but in the so-called hot CNO cycle. It is characterized by the fact that in the branching between the slow beta-decay of $^{13}$N and a further proton capture to $^{14}$O, the proton capture wins because of a highly enhanced reaction rate for such conditions. In a similar way hot CNO-type cycles for elements beyond Ne re-arrange nuclei up to Mg and Si. When including ejecta of such nova explosions, a few not highly abundant isotopes up to Mg and Si can be considered to have a non-negligible contribution from novae \citep{Jose:2016}.
 
 In type~I X-ray bursts, due to the explosive ignition of H- and subsequent He-burning reactions, matter from the hot CNO cycle can even be transferred to heavier nuclei up to $^{100}$Sn via the rp-process. 
 This is an important stellar explosion, observed in X-rays, but the explosion energy is likely not sufficient to eject matter out of the high gravitational binding of the neutron star. A detailed discussion of novae and X-ray bursts can be found in \cite{Jose:2016}, the evolution of knowledge on the nuclear reactions evolved over many years from \cite{Wiescher.ea:1986} over \cite{Schatz.ea:1998} to \cite{Cyburt.ea:2010,Schatz.Ong:2017,Meisel.ea:2020}. 

 In addition to the above mentioned explosive burning processes, there exist other processes producing some isotopes beyond iron that are not accessible by neutron capture processes. They include the p-process or $\gamma$-process (occurring as a byproduct of explosive Ne-burning which was discussed before) and the $\nu$p-process (Fig.~\ref{fig:nuc_chart}), responsible for proton-rich stable nuclei up to $A=80$--$90$.
The p-process consist mainly of photo dissociation of existing heavy isotopes. This moves matter from nuclei previously produced by the s- and r-process to the proton-rich side of stability. The initial suggestion was that conditions in a hydrodynamic shock wave triggers this process when running through layers of an exploding star that contain already heavy nuclei from previous generations. For an overview of possible sources for p-rich nuclei and the relevant nuclear physics, see \cite{rauscher13,Roberti.Pignatari.ea:2023}.

An additional option to produce light p-nuclei is the so-called $\nu$p-process in proton-rich ejecta in supernova explosions \citep{Froehlich.Martinez-Pinedo.ea:2006,Wanajo.Janka.ea:2011,Ghosh.ea:2022}.

\subsection{What type of reactions are needed in explosive burning?}

\begin{figure}[h!]
\includegraphics[width=\textwidth,angle=0]{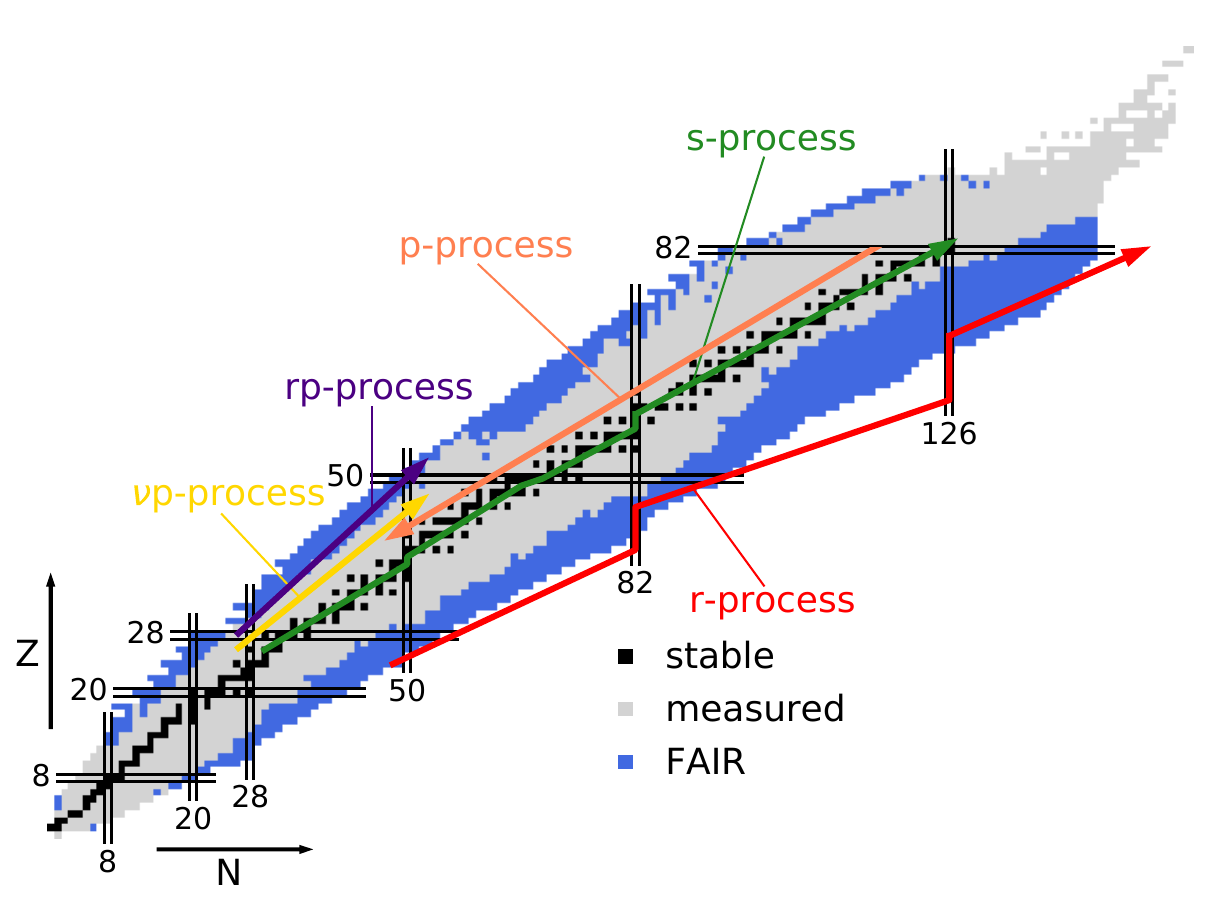}
\caption{(courtesy of M. Jacobi and A. Arcones) The nuclear chart with stable isotopes marked with black boxes, the gray region indicates nuclei that have been produced already in the laboratory, and the light blue region shows the exotic isotopes that rare isotope facilities will discover. The various explosive nucleosynthesis processes discussed in previous subsections are schematically indicated by color lines. Explosive C, Ne, O, and Si burning (unless happening under very neutron-rich conditions) involve stable and unstable nuclei close to stability. The s-process, taking place in hydrostatic He-burning, is also shown here, but it can to a large extent (with exceptions) be investigated with experimentally determined neutron capture cross sections of stable nuclei.}
\label{fig:nuc_chart}
\end{figure}

After going through the various hydrostatic and explosive burning processes which need input for nuclear reaction cross sections, we attempt to give a short overview on the regions in the nuclear chart where such reaction cross sections and reaction rates are needed. A summary of all these processes and their actions in specific regions of the nuclear chart is given in Fig.\ref{fig:nuc_chart}. These include proton capture reaction in the rp-process and $\nu$p-process, photodisintegration reactions in the p- or $\gamma$-process, and neutron capture reactions in the r-process. The other explosive burning processes mentioned in the previous subsections work closer to stability, but also include to some extend unstable nuclei, for which charged particle and neutron capture reactions need to be predicted.

\def\cm{~{\rm cm}}
\def\gcm{~{\rm g~cm}^{-3}}
\def\kmsec{~{{\rm km~sec}^{-1}}}
\def\ergsec{~{\rm erg~sec}^{-1}}
\def\a{\alpha}
\def\b{\beta}
\def\g{\gamma}
\def\G{\Gamma}
\def\d{\delta}
\def\D{\Delta}
\def\e{\epsilon}
\def\vare{\varepsilon}
\def\z{\zeta}
\def\th{\theta}
\def\TH{\Theta}
\def\varth{\vartheta}
\def\i{\iota}
\def\k{\kappa}
\def\l{\lambda}
\def\L{\Lambda}
\def\r{\rho}
\def\varr{\varrho}
\def\s{\sigma}
\def\vars{\varsigma}
\def\t{\tau}
\def\u{\upsilon}
\def\oo{\omega}
\def\O{\Omega}
\def\na{\nabla}
\def\pa{\partial}
\def\p{\prime}
\def\pp{\prime\prime}
\def\sm{~{\rm M}_{\odot}} 
\def\cl{\centerline}
\def\mkt{^{\mu/kT}}

\section
{Thermonuclear Rates and the Hauser-Feshbach Formlism}

In the section on hydrostatic burning stages, we indicated that at least all the early burning stages
rely on experimentally determined nuclear reaction cross sections.
In late burning stages and explosive burning a number of unstable nuclei are produced and a
much larger amount of stable nuclei. It is thus advisable to look into
specific experimental techniques or alternatively into reliable
theoretical approaches which can provide the necessary information.
Explosive burning in supernovae involves in general intermediate mass
and heavy nuclei. Due to a large nucleon number they have intrinsically
a high density of excited states. A high level density in the 
compound nucleus at the appropriate excitation energy allows to
make use of the statistical model approach for compound nuclear 
reactions (Hauser-Feshbach), which averages over resonances. Therefore, it is often colloquially termed that
the statistical model is only applicable for intermediate and heavy
nuclei, which gives repeatedly rise
to misunderstandings. The only necessary condition for application
of the statistical model formalism is a large number of resonances
at the appropriate bombarding energies, so that the cross section can be
described by average transmission coefficients. Whether we have a light or
heavy nucleus is only of secondary importance. 
A high level density automatically implies that the
nucleus can equilibrate in the classical compound nucleus picture.

As the capture of an alpha particle leads usually to larger Q-values than
neutron or proton captures, the compound nucleus is created at a higher
excitation energy and especially in the case of alpha captures it
is often even possible to apply the Hauser-Feshbach formalism for nuclei
as light as Li. Another advantage of alpha capture 
is that the
capture Q-values vary very little with the N/Z-ratio of a nucleus.
This means that the requirements are equally well fulfilled
near stability as at the proton or neutron drip lines, which makes 
the application very safe for all intermediate and heavy nuclei
(and for light nuclei the level density test has to be done 
individually). Opposite to the behavior for alpha-induced reactions,
the reaction Q-values for proton or neutron captures vary strongly
with the N/Z-ratio, leading eventually to vanishing Q-values at
the proton or neutron drip line. For small Q-values the compound
nucleus is created at low excitation energies and also for intermediate
nuclei the level density can be quite small. Therefore, it is not
advisable to apply the statistical model approach close to the 
proton drip line for intermediate nuclei.
For neutron captures close to the neutron drip line in r-process
applications it might be still permissible for heavy and often
deformed nuclei, which have a high level density already at
very low excitation energies. On the other hand, this method is not
applicable for neutron-rich light nuclei. Where predictions for nuclear reaction rates
are needed to perform investigations in the related astrophysical processes has been 
outlined in Fig.\ref{fig:nuc_chart}.

The statistical model approach has been employed
in the calculation of thermonuclear reaction rates for astrophysical 
purposes by many researchers, starting with \cite{Truran.Hansen.ea:1966}, who only
made use of ground state properties. \cite{Arnould:1972} pointed out the
importance of excited states. Extended compilations have been provided \citep[][for updates see \footnote{https://nucastro.org, https://reaclib.jinaweb.org and https://www-nds.iaea.org, including TALYS results}]{Holmes.Woosley.ea:1976,Woosley.Fowler.ea:1978,Thielemann.Arnould.Truran:1986,Rauscher.Thielemann:2000,Goriely.Hilaire.Koning:2008}. The codes for the latter three entries are known under the 
names SMOKER, NON-SMOKER, and TALYS. More recently, the code SMARAGD
\citep{rauscher:smaragd,Rauscher:2011} has been developed as a successor to
NON-SMOKER, with improvements in the treatment of charged particle and photon transmissions, in
the nuclear level density, and with inclusion of recent experimental information on
low-lying excited states. Reaction rate compilations obtained from calculations using
NON-SMOKER and TALYS
are presently the ones utilized
in large scale applications in all subfields of nuclear astrophysics,
when experimental information is unavailable. SMARAGD is being used to analyze
experimental data and for improvements of smaller sets of reaction rates \citep{Cyburt.ea:2010} but
a large-scale set of reaction rates has not been published yet.

A high level density in the compound nucleus permits to use averaged
transmission coefficients $T$, which do not reflect a resonance behavior,
but rather describe absorption via an imaginary part in the (optical)
nucleon-nucleus potential \citep[for details see, e.g.][]{Mahaux.Weidenmueller:1979}. This leads to the well known expression

\begin{align}
\sigma^{\mu \nu}_{i} (j,o;E_{ij})= &
{{\pi \hbar^2 /(2 \mu_{ij} E_{ij})} \over {(2J^\mu_i+1)(2J_j+1)}} \label{eq:HFmu}\\
& \times \sum_{J,\pi} (2J+1){{T^\mu_j (E,J,\pi ,E^\mu_i,J^\mu_i,
\pi^\mu_i) T^\nu_o (E,J,\pi,E^\nu_m,J^\nu_m,\pi^\nu_m)} \over
{T_{tot} (E,J,\pi)}}\nonumber
\end{align}

for the reaction $i^\mu (j,o) m^\nu$ from the target
state $i^{\mu}$ to the exited state $m^{\nu}$ of the final nucleus, with
center of mass energy E$_{ij}$ and reduced mass $\mu _{ij}$. $J$ denotes the
spin, $E$ the excitation energy, and $\pi$ the parity of excited states.
When these properties are used  without subscripts they describe the compound
nucleus, subscripts refer to the participating nuclei in the
reaction $i^\mu (j,o) m^\nu$
and superscripts indicate the specific excited states.
Experiments measure $\sum_{\nu} \sigma_{i} ^{0\nu} (j,o;E_{ij})$,
summed over all excited states of
the final nucleus, with the target in the ground state. Target states $\mu$ in
an astrophysical plasma are thermally populated and the astrophysical cross
section $\sigma^*_{i}(j,o)$ is given by
\begin{figure}
    \centering
    \includegraphics[width=0.8\linewidth]{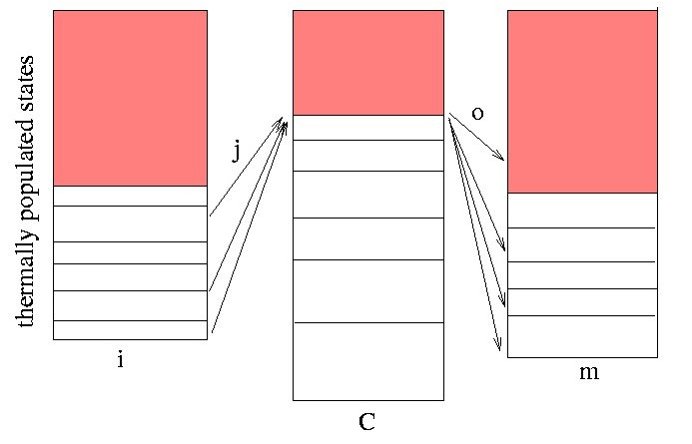}
    \caption{Energetics for the reaction $i(j,o)m$ and its inverse $m(o,j)i$
for transitions through a resonance state $(E,J^\pi)$ in the compound
nucleus $C$.}
    \label{fig:compound}
\end{figure}

\begin{equation}
\sigma^*_{i} (j,o;E_{ij}) = {\sum_\mu (2J^\mu_i+1) \exp(-E^\mu_i /kT)
\sum_\nu \sigma^{\mu \nu}_{i}(j,o;E_{ij}) \over \sum_\mu (2J^\mu_i+1)
 \exp(-E^\mu_i/kT)}.
 \label{eq:HF*}
\end{equation}

The summation over $\nu$ replaces $T_o^{\nu}(E,J,\pi)$ in Eq.(\ref{eq:HFmu}) by
the total transmission coefficient

\begin{align}
T_o (E,J,\pi) = &\sum^{\nu_m}_{\nu =0} 
T^\nu_{om}(E,J,\pi,E^\nu_m,J^\nu_m, \pi^\nu_m) \\
&+ \int\limits_{E^{\nu_m}_m}^{E-S_{m,o}} \sum_{J_m,\pi_m}
T_{om}(E,J,\pi,E_m,J_m,\pi_m)\rho(E_m,J_m,\pi_m) dE_m.\nonumber
\end{align}

Here $S_{m,o}$ is the channel separation energy, and the summation over
excited 
states above the highest experimentally
known state $\nu_m$ is changed to an integration over the level density
$\rho$.
The summation over target states $\mu$ in Eq.(\ref{eq:HF*}) has to be generalized
accordingly. 

The important ingredients of statistical model calculations
are the particle and $\gamma$-transmission coefficients $T$ and
the level density of excited states $\rho$. Therefore, the reliability of the 
of such calculations is determined by the accuracy with which these components 
can be evaluated. In the following we want to discuss
the methods utilized to estimate these quantities and recent improvements.

\subsection{Particle Transmission Coefficients}

The transition from an excited state in the compound nucleus $(E,J,\pi)$
to the state $(E^\mu_i,J^\mu_i,\pi^\mu_i)$ in nucleus $i$ via the emission of
a particle $j$ is given by a summation over all quantum mechanically allowed
partial waves

\begin{equation}
T^\mu_j (E,J,\pi,E^\mu_i,J^\mu_i,\pi^\mu_i) =
\sum_{l=\vert J-s \vert}^{J+s}
\sum_{s=\vert J^\mu_i -J_j \vert}^{J^\mu_i + J_j}
T_{j_{ls}} (E^\mu_{ij}).  \label{partT}
\end{equation}

\begin{figure}
    \centering
    \includegraphics[angle=180.5,width=0.6\linewidth]{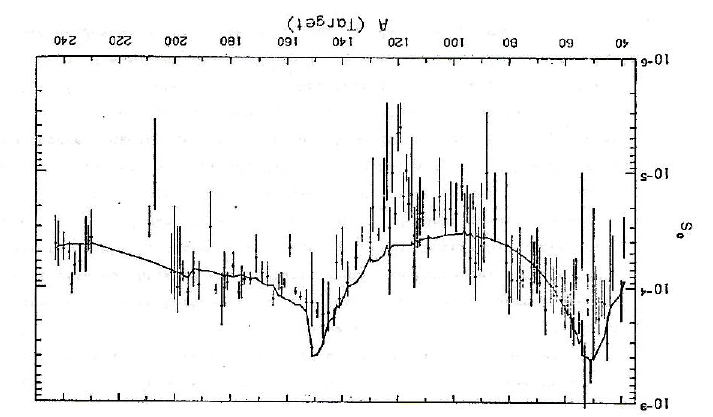}
    \caption{
Neutron strength function when utilizing a sophisticated optical potential \citep{Jeukenne.Lejeune.Mahaux:1977}.
An equivalent square well potential would result in
a straight line. When utilizing a spherical potential a deviation from experiment around A=160 would 
show up, with an $S_o$-line similar to a sinus curve. This is avoided here by 
treating deformed nuclei in a very simplified way (without utilizing more sophisticated coupled
channel calculations) by using an effective spherical potential of equal
volume, based on averaging the deformed potential over all possible
angles between the incoming particle and the orientation of the deformed
nucleus. Image reproduced with permission from \cite{Cowan.Thielemann.Truran:1991}, copyright by Elsevier.}
    \label{fig:neutronstrength}
\end{figure}

Here the angular momentum $\vec l$ and the channel spin $\vec s =\vec J_j+
\vec J^\mu_i$ couple to $\vec J = \vec l +\vec s$. The individual transmission
coefficients $T_l$ are calculated by solving the Schr\"odinger equation
with an optical potential for the particle-nucleus interaction. While all early
studies of thermonuclear reaction rates \citep{Truran.Hansen.ea:1966,Arnould:1972,Holmes.Woosley.ea:1976,Woosley.Fowler.ea:1978}
employed optical square well
potentials and made use of the black nucleus approximation, the other compilations utilize optical potentials from  microscopic
calculations. The resulting s-wave neutron strength
function (the average ratio of the width of states populated by s-wave neutrons and the appropriate level spacing $D$ for a 1 eV capture energy) can be expressed by $S_0=<\Gamma^o/D> \vert_{\rm 1eV}=(1/2\pi )T_{n(l=0)}(\rm 1eV)$. A specific
optical potential choice \citep{Jeukenne.Lejeune.Mahaux:1977} with corrections for the imaginary part and updated parameters for low energies
\citep{lejeune:1980} is shown in Fig.\ref{fig:neutronstrength} (utilized in the NON-SMOKER code).

In a similar way there exist a variety of optical potentials for protons \citep[including][with the appropriate energetics, see also many more options in the TALYS package]{Jeukenne.Lejeune.Mahaux:1977}.
Alpha particles have been treated in many cases with a
phenomenological Woods-Saxon potential based on extensive data by McFadden and Satchler \citep{McFadden.Satchler:1966}.
In general, for alpha particles and heavier projectiles,
the best results can probably be obtained with folding potentials \cite[see e.g.][]{Satchler.Love:1979,Mohr.Fulop.ea:2020}.

\subsection{$\gamma$-Transmission Coefficients}

At low energies photon captures on nuclei lead to relatively simple excited
states, often involving only few particles in the nucleus. Beyond about 10 MeV
so-called giant resonances are populated which correspond to a collective motion involving many if not all
particles of a nucleus. They occur as electric or magnetic multipole resonances and can be excited by 
capture of a photon or can decay by emitting a photon.
The dipole E1 and M1 resonances dominate. In a macroscopic way they can be understood by protons
oscillating against neutrons (a) in the Goldhaber-Teller mode where fixed proton and neutron density distributions move
against each other or (b) in the Steinwedel-Jensen mode with density oscillations of neutrons against protons taking place 
within a fixed nuclear surface. In such "macroscopic" descriptions the resonance energy and its width can be explained with
the ingredients of macroscopic-microscopic nuclear mass models \citep[e.g.][]{Thielemann.Arnould.Truran:1986,Thielemann.Arnould:1983,Rauscher.Thielemann:2000}.
Extended collections of experimental data, showing essentially a Lorentzian distribution as a function of energy,
have been provided in the past \citep[e.g.][]{Berman.Fultz:1975,Junghans.ea:2008}, see for recent comprehensive reviews \citep{Ishkhanov.Kapitonov:2021,Liang.Litvinova:2022}, the latter reference corresponding to the chapter by Liang \& Litvinova in this Handbook, containing a detailed derivation within a microscopic (rather than phenomenological "macroscopic") description of these giant resonances. Both types of options to choose are available in the TALYS code.
In the more phenomenological approach the E1 transitions
are calculated on the basis of the Lorentzian representation of the
Giant Dipole Resonance (GDR). Within this model, the E1 transmission
coefficient
for the transition emitting a photon of energy $E_{\gamma}$ in a nucleus
$^A_N Z$ is given by

\begin{equation}
T_{E1}(E_{\gamma}) = {8 \over 3} {NZ \over A} {e^2 \over \hbar c}
{ {1+\chi}
\over mc^2} \sum_{i=1}^2 {i \over 3} { {\Gamma_{G,i}^2 E^4_\gamma} \over
{(E_\gamma^2 -E^2_{G,i})^2 + \Gamma^2_{G,i} E^2_\gamma}}.
\label{gammaT}
\end{equation}

\begin{figure}
    \centering
    \includegraphics[width=0.4\linewidth]{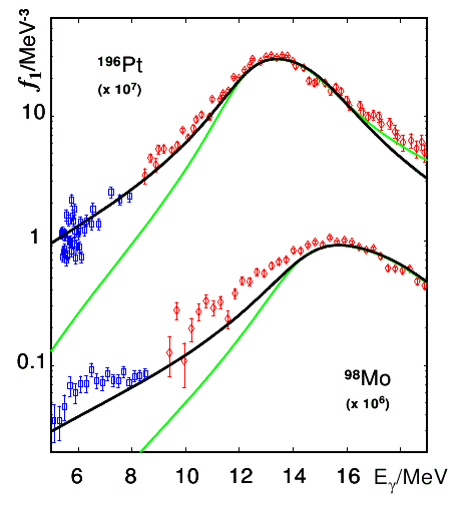}
     \includegraphics[width=0.4\linewidth]{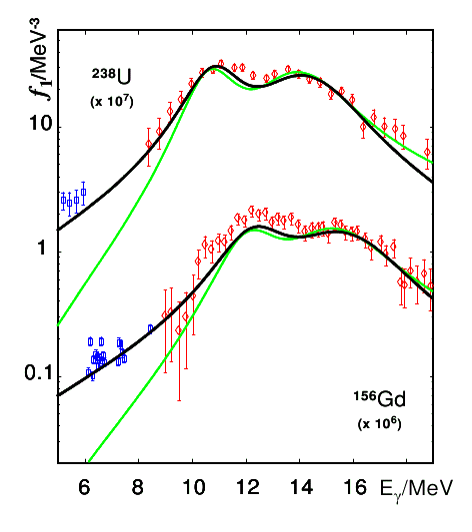}
    \caption{Experimental strength functions of the Giant Dipole Resonance $f_{E1}(E_\gamma)=(T_{E1}(E_\gamma)/2\pi) (E_\gamma)^{-3}$ for spherical and deformed nuclei from \cite{Junghans.deJong.ea:1998}. For deformed nuclei the Lorentzian form splits into two sub-resonances (see text). The lines represent two choices for the description of the GDR width chosen by \cite{Junghans.ea:2008}. This includes, however, also low-lying E1 strength, which will be discussed below in more detail. Image reproduced with permission from \cite{Junghans.ea:2008}, copyright by Elsevier.
}
    \label{fig:Junghans}
\end{figure}

Here $\chi(=0.2)$ accounts for the neutron-proton exchange contribution and the
summation over $i$ includes two terms which correspond to the split of the
GDR in statically deformed nuclei, with oscillations along (i=1) and
perpendicular (i=2) to the axis of rotational symmetry. Specific experimental results \citep{Junghans.ea:2008}
are shown in Fig.\ref{fig:Junghans}.

As discussed above, the GDR resonance energy $E_G$ can be predicted within macroscopic (hydrodynamic) approaches
based on the ingredients of marcoscopic-microscopic mass models \citep{Thielemann.Arnould:1983,Rauscher.Thielemann:2000}, which gives excellent fits to the GDR energies and
can also predict the split of the resonance for deformed nuclei, when
making use of the deformation. In that case, the two resonance energies are related to the
mean value by the following expression \citep{Danos:1958}

\begin{align}
E_{G,1}+2E_{G,2}&=3E_G \nonumber\\
E_{G,2}/E_{G,1}&=0.911 \eta +0.089. 
\label{giantR}
\end{align}

\begin{figure}
    \centering
    \includegraphics[width=0.95\linewidth]{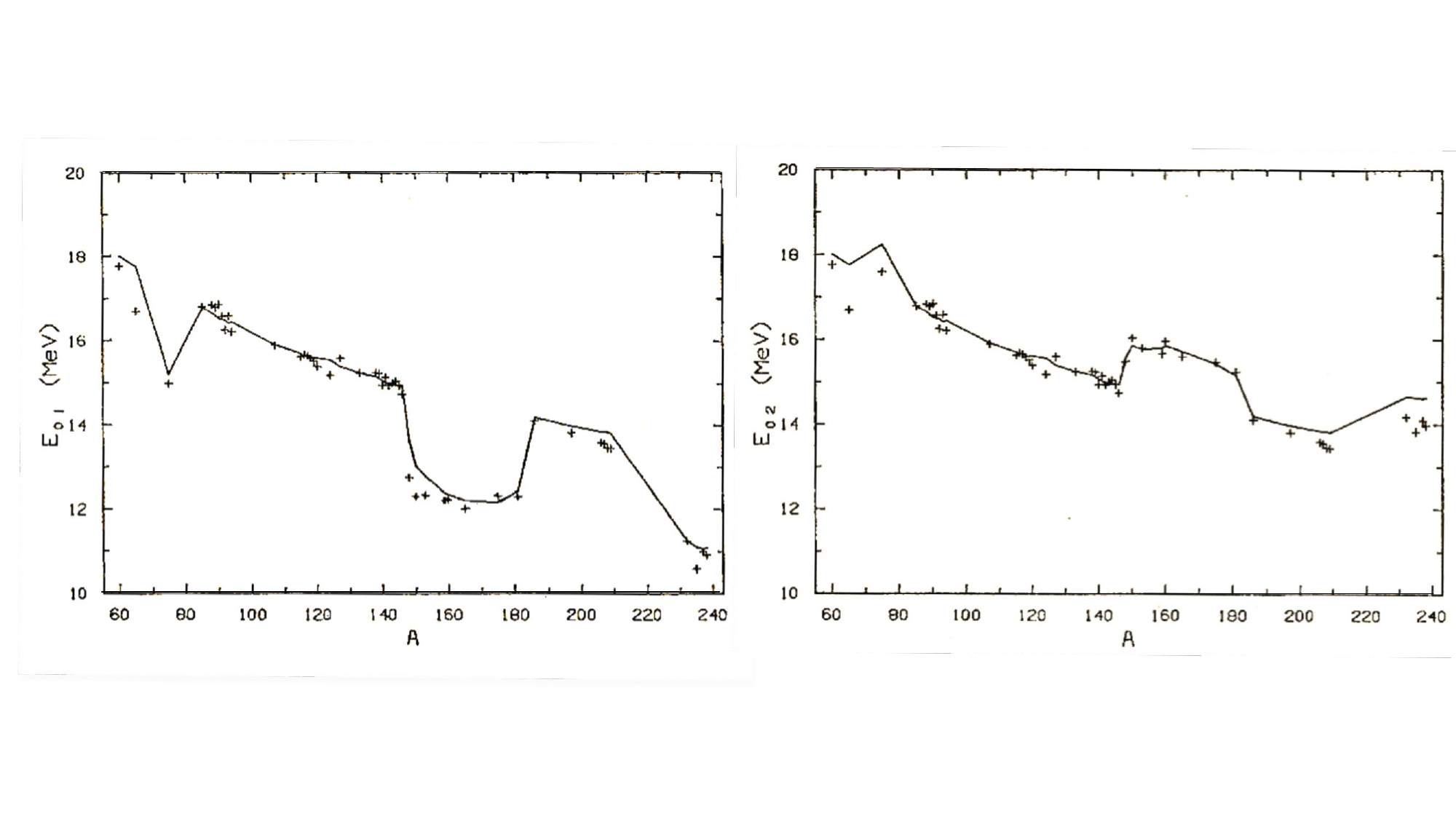}
    \caption{E1 giant dipole resonance energy for oscillations parallel ($E_{01}$) and perpendicular ($E_{02}$) to the rotational symmetry of deformed nuclei. For spherical nuclei both energies coincide. The energies are predicted with parameters from microscopic-macroscopic mass models (as utilized in the NON-SMOKER code), deformations taken from experiments or mass models as well.Image reproduced with permission from \cite{Cowan.Thielemann.Truran:1991}, copyright by Elsevier}. 
    \label{fig:egdr}
\end{figure}

$\eta$ is the ratio of (1) the diameter along the nuclear
symmetry axis and (2) the diameter perpendicular to it. It can be
obtained from the experimentally known deformation or mass model
predictions (see Fig.\ref{fig:egdr}).

The width of the GDR $\Gamma_G$ is less understood, but can also predicted in a phenomenological approach via the ingredients of macroscopic-microscopic nuclear mass models which satisfactorily reproduces the experimental data for spherical
and deformed nuclei. The
GDR width can be described as a superposition  of a macroscopic width due
to the viscosity of the nuclear fluid and a coupling to quadrupole surface
vibrations of the nucleus \citep{Thielemann.Arnould:1983,Rauscher.Thielemann:2000,Goriely:1998}. 
The width is reduced near magic nuclei, because these closed shell nuclei
act within the hydrodynamic picture in a stiff way and couple less to the quadrupole surface vibrations.
Alternatively microscopic approaches have been utilized within the quasi-particle Random Phase Approximation (QRPA) (see e.g., besides a macroscopic approach similar to the above description, different options in the TALYS code based on Skyrme Hatree-Fock BCS, Skyrme Hartree-Fock-Bogoluibov, the temperature-dependent relatistic mean field or the Gogny Hartree-Fock-Bogoluibov model) or also the Relativistic Quasiparticle Time Blocking approximation (RQTBA) framework \citep{Litvinova.Loens.ea:2009}. For a general review see \cite{Liang.Litvinova:2022}.

The total radiation width at the neutron separation energy for nuclei in experimental
compilations can be calculated via the total photon transmission
coefficients for E1 radiation from a compound nucleus state with energy
$E$, spin $J$, and parity $\pi$ is given by

\begin{align}
T_\gamma (E,J,\pi) = &\sum^{\nu_C}_{\nu =0} 
T^\nu_\gamma(E,J,\pi,E^\nu_C,J^\nu_C,\pi^\nu_C)\nonumber\\ 
&+ \int\limits_{E^{\nu_C}_m}^E \sum_{J_C,\pi_C}
T_\gamma (E,J,\pi,E_C,J_C,\pi_C)\rho(E_C,J_C,\pi_C) dE_C, 
\label{Tgamma}
\end{align}

where the first term represents a summation over the known low-lying
states $\nu$ up to an excitation energy $E_\omega=E_c^{\nu_c}$, while an
integration involving the level density $\rho$ is performed at higher
excitation. $T^\nu_\gamma(E,J,\pi,E^\nu_C,J^\nu_C,\pi^\nu_C)$
is either zero if the E1 selection rules are violated, or equal to 
$T_{E1}(E_\gamma=E-E^\nu_C)$ otherwise. With $T_{E1}$  from Eq.(\ref{gammaT}) 
the average radiation width $<\Gamma_\gamma(E,J,\pi)>$ (not to be confused with the GDR width) is related to
the total photon transmission coefficient by $<\Gamma_\gamma(E,J,\pi)>=
T_\gamma(E,J,\pi)/(2\pi \rho(E,J,\pi))$. This way the nuclear level density (which will be discussed in the 
next subsection) enters in the
average radiation width. Extensive sets of
measured average radiation widths come from thermal s-wave ($l=0$) neutron
capture, the relevant theoretical
quantity to be compared with such experimental data is

\begin{equation}
<\Gamma_\gamma>_0 =
{J_i+1 \over  2J_i+1} <\Gamma_\gamma(S_n,J_i+1/2,\pi_i)>+
{J_i \over  2J_i+1} <\Gamma_\gamma(S_n,J_i-1/2,\pi_i)>, 
\label{eq:radwidth}
\end{equation}

where $S_n$ is the compound nucleus neutron separation energy and
$J_i(\pi_i)$ is the spin (parity) of the target nucleus. Utilizing the
methods outlined above within a macroscopic-microscopic approach for the Lorentzian
form of the E1 giant dipole resonance leads to a global agreement generally within
a factor of 1.5 for nuclei experimentally accessible. This involves a correction to an energy-dependent width 
due to the fact, that for low energy gamma-transitions which are part of the Giant resonance, the Lorentz curve
is suppressed and, in addition, the GDR width increases with excitation energy.
This led  to $\Gamma_G(E_\gamma)=\Gamma_G\times(E_\gamma/E_G)^{\delta}$ with $\delta$=0.5
\citep{McCullagh.ea:1981}, for a slightly different term (close to $\delta=1$) have a look at \cite{Goriely.Plujko:2019}.

\begin{figure}[h!]
    \centering
    \includegraphics[width=\textwidth]{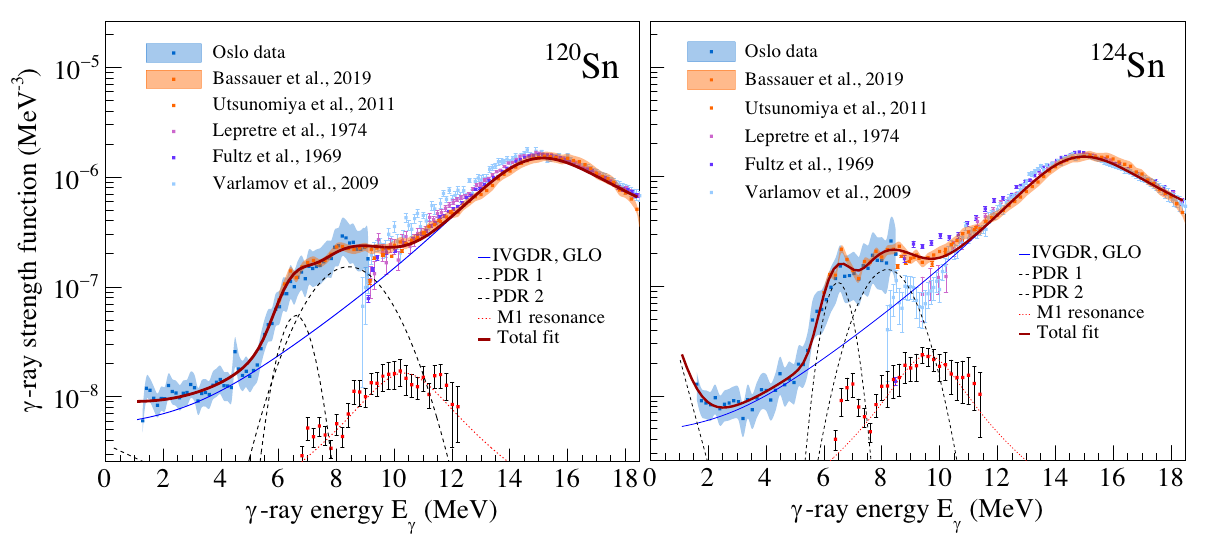}
    \caption{(courtesy of M. Markova) Total gamma ray strength function as a function of transition energy (see also Fig.\ref{fig:Junghans}) for two Sn isotopes. See especially the contribution of the low-lying strength in the E1 PDR, the two M1 spin flip and scissor mode contributions, and the upbend at lowest energies.}
    \label{fig:markova}
\end{figure}

An additional entity of the low-lying E1 response below the GDR energy
was experimentally noticed \citep[see e.g.][termed as Pigmy dipole resonance PDR]{Goriely:1998} and led other authors \citep{Junghans.ea:2008} to an artificial broadening of the GDR with a $\delta$=1.5 (see the broader curve in Fig.\ref{fig:Junghans}). Microscopically it can be well described by Skyrme+QRPA or RMF+cQRPA approaches, macroscopically it is interpreted as an oscillation of the neutron skin against the $N=Z$ core. Further investigations into this topic have been undertaken theoretically and experimentally \citep[e.g][]{Litvinova.Loens.ea:2009,Guttormsen.ea:2022}. An approximate description for the E1 PDR within a phenomenological macroscopic approach, based on the neutron skin thickness, has been provided \citep{VanIsaker.ea:1992} and utilized \citep{Goriely:1998}. Recent developments include an advanced understanding of the PDR (with E1 and M1 contributions)
and its total effect on capture cross sections \citep{Markova.ea:2021,Guttormsen.ea:2022} as shown in Fig.\ref{fig:markova}. 

Various treatments are available for the M1 giant resonance in a similar functional form as discussed above in Eq.(\ref{gammaT}), but with typically smaller resonance energies and widths and a smaller normalisation factor (see the M1 component in Fig.\ref{fig:markova}).
This involves two M1 modes, the spin-flip and the scissor mode (the latter only present in deformed nuclei), for which empirical expressions have been derived in \cite{Goriely.Plujko:2019} in forms identical to Eq.(\ref{Tgamma}), but with different scaling factors and typical resonance energies around 7-10 MeV, respective 2.5-4 MeV and corresponding widths of about 4 respective 1.5 MeV \citep[for shell model based predictions see e.g.][]{Loens.Langanke.ea:2008,Sieja:2018}. An additional effect relates to the fact that for small energies the Brink hypothesis (the transition strength as a function of energy is independent of the excitation energy of the state) is apparently broken at low energies, leading to a different behavior for absorption in comparison to de-excitation. This results, based on shell model calculations, in a constant limit towards lowest energies for the E1 strength and an upbend in case of the M1 strength  \citep{Goriely.Hilaire.ea:2018,Goriely.Plujko:2019}.
The effects can nicely be seen in Fig.\ref{fig:markova}, with the upbend at lowest energies, the two M1 modes, and an extended E1 PDR. 
While the low-lying E1 strength contributes only on the less than 10\% level close to stability, for very neutron-rich nuclei with an extended neutron skin it can lead to large enhancements and the low-lying M1 strength and especially the upband at lowest energies can have a quite sizable effect close to the neutron drip line.

Another effect has to be taken into account for alpha-capture reactions on self-conjugate ($N=Z$) nuclei, because due to isospin selection rules $\gamma$-transitions between isospin $I=0$ levels are forbidden. This leads to a suppression of corresponding 
alpha-capture cross sections which can be treated via a suppression of the appropriate $\gamma$-transitions \citep{Rauscher.Thielemann.Goerres.Wiescher:2000}.

\subsection{Fission}

Along the valley of stability, actinide and transuranic
nuclei 
become increasingly unstable as a result of fission in their ground states.
This loss of  stability arises because the disruptive Coulomb force, which
increases as $Z^2$, overcomes the cohesive surface tension, 
proportional to $A^{2/3}$.
The nuclear mass, i.e. potential
energy for a given
$(Z,A)$, as a function of deformation, has a local minimum at the ground state
deformation. It  first increases with increasing deformation until reaching
a maximum at the saddle point, from which point onward fission is inevitable.
This potential barrier (=fission barrier) can have very small tunneling
transmission coefficients and therefore very long fission half-lives.
While the simple liquid-drop approach to nuclear masses leads only to a single barrier plotted as a function of deformation, the barrier is in most cases split into a double-humped (or even multiple) barrier prescription,
due to shell effects \citep{Strutinsky:1967,Strutinsky:1968}. The higher of the peaks is
denoted the fission barrier, since states beyond that excitation energy can
actually fission instantaneously.

\begin{figure}[h!]
    \centering
    \includegraphics[width=0.7\textwidth]{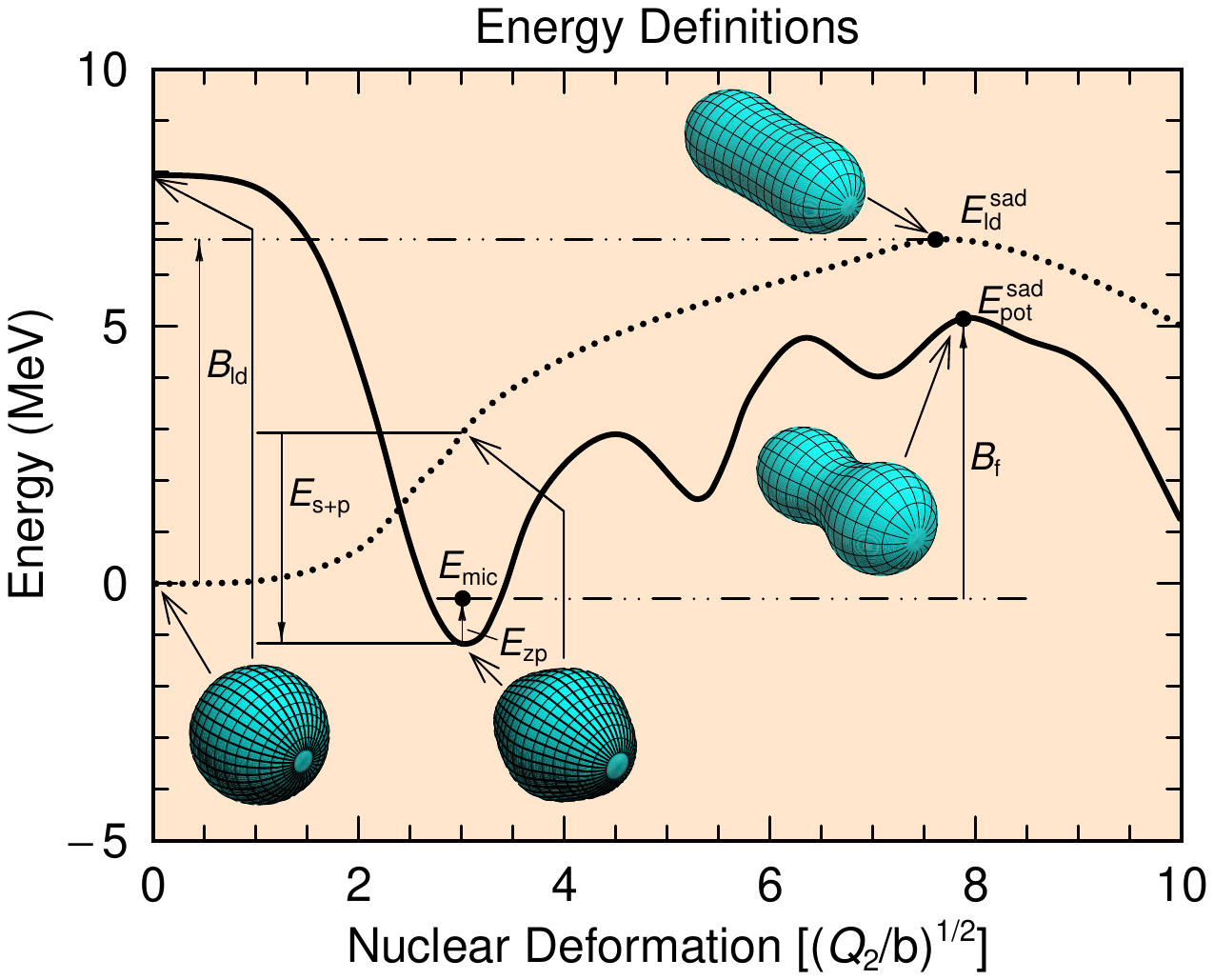}
    \includegraphics[width=0.6\textwidth]{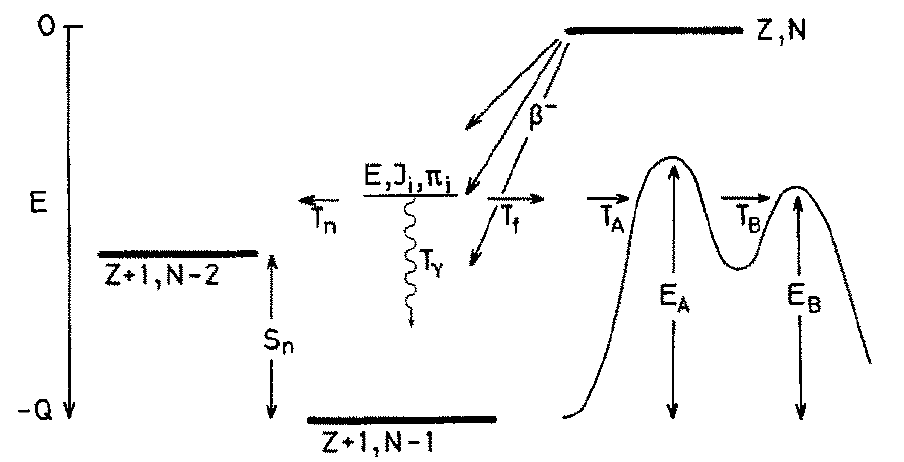}
    \caption{top: The Energy (or respective mass of the nucleus) as a function of the quadrupole moment $Q_2$, which measures the elongation of the nucleus due to deformation, based on macroscopic-microscopic fission potential-energy calculations for $^{232}$Th \citep{Moller.Sierk.ea:2015}. The
dotted line corresponds to the macroscopic “liquid-drop” energy along a specified path; the solid line is the total macroscopic-microscopic
energy along a sequence of different shapes. $Q_2$ = 0 stands for a spherical shape. From the ground state towards larger deformations the
total-energy curve is given along the optimal fission path that includes all minima and saddle points, identified along this path in a
five-dimensional deformation space (elongation, neck diameter, left-fragment spheroidal
deformation, right-fragment spheroidal deformation, and nascent-fragment mass asymmetry). bottom: similar plot of the potential energy as a function of deformation with a double-humped fission barrier, as the typical feature for actinide nuclei \citep{Thielemann.Metzinger.Klapdor:1983}. The energetics are shown here for an application for beta-delayed fission. [left] Image provided from the database of P. M\"oller, see \url{https://t2.lanl.gov/nis/molleretal/}, [right] image reproduced with permission from \cite{Thielemann.Metzinger.Klapdor:1983}, copyright by Springer Nature.}
    \label{fig:Mollerfisbarr}
\end{figure}
\begin{figure}[h!]
    \centering
    \includegraphics[width=0.9\textwidth]{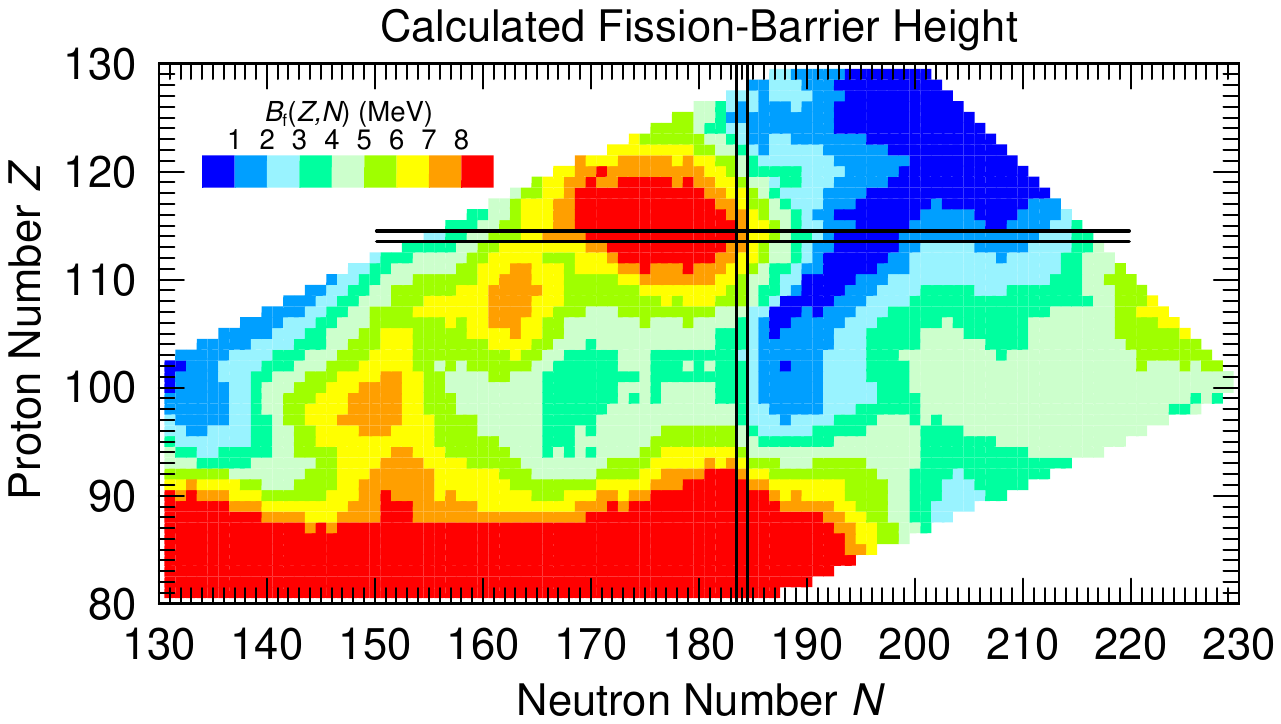}
    \caption{Fission barrier heights $B_f$  based on macroscopic-microscopic fission potential-energy calculations \citep{Moller.Sierk.ea:2015}. Image provided from the database of P. M\"oller, see \url{https://t2.lanl.gov/nis/molleretal/}.}
    \label{fig:Mollerfisbarr2}
\end{figure}

 
More generally, the mass (potential energy) can be evaluated not only along
a deformation path but rather as a function of several deformation
parameters. One possibility is the choice of utilizing the parameters elongation, neck diameter, left-fragment spheroidal
deformation, right-fragment spheroidal deformation, and nascent-fragment mass asymmetry. Fig. \ref{fig:Mollerfisbarr} (top panel) shows such a
potential energy contour plot for $^{232}$Th \citep{Moller.Sierk.ea:2015}.
Because the potential energy (=mass) as a function of deformation must be
calculated for a particular choice of mass model, many theoretical efforts have been undertaken \citep[see e.g.][and references therein]{Schunk.Regnier:2022} and the results will be
dependent on these mass models. Fig.\ref{fig:Mollerfisbarr2} gives the results of \cite{Moller.Sierk.ea:2015}.
Investigations utilized in astrophysical applications developed from 1980 until present \citep[see e.g.][]{Howard.Moeller:1980,Myers.Swiatecki:1999,Mamdouh.Pearson.ea:2001,Goriely.Hilaire.ea:2009,Vassh.Vogt.ea:2019,Giuliani.Martinez.ea:2020}.

For most of the actinides the fission barrier development as a function of deformation can be described within the so-called double-humped fission barrier \citep{Bjornholm.Lynn:1980} and the fission probability can be calculated within the complete-damping formalism, making use of two Hill-Wheeler inverted parabola barriers \citep{Lynn.Back:1974}, as utilized in a number of applications \citep{Thielemann.Metzinger.Klapdor:1983,Panov.Kolbe.ea:2005,Panov.Korneev.ea:2010} (see Fig.\ref{fig:Mollerfisbarr}, bottom panel), otherwise a more complex quantum mechanical approach for the penetration through the barrier potential has to be performed \citep{Goriely.Hilaire.ea:2009}.

As discussed previously, excited states close to or above the fission barrier
will
have much shorter fission half-lives than ground states, and therefore
processes
like neutron capture or beta-decay, which populate excited states of the
compound nucleus or daughter nucleus, will lead to increased probabilities for 
fission relative to the
ground state. Therefore, neutron-induced and beta-delayed fission can play
an important role in astrophysics, including the  
the r-process.  

The statistical model calculations discussed in this section can be
extended to include a fission channel,
in order to calculate cross sections for neutron-induced fission.
The cross section for a reaction $i^0(j,o)m$, from the target ground
state $i^0$ to all excited states $m^{\nu}$ of a final nucleus
with center of mass energy E$_{ij}$ and reduced mass $\mu _{ij}$,
is again given by Eqs.(\ref{eq:HF*}), with the summation taken over all final states
$\nu$.
Therefore, the sum of $T_o^\nu$ is again replaced by the expression
for $T_o$. The ratio going into the fission channel is then $T_o/T_{tot}$
with $o=f$, {\it i.e.} the outgoing channel is the fission channel.
When making use of a double-humped fission barrier, or a more complex transmission calculations, the result and can be expressed
in terms of the fission probability
$P_f(E,J,\pi )= T_f(E,J,\pi )/T_{tot}(E,J,\pi)$.
With a value obtained for $P_f$ and the calculated transmission coefficients
$T_o$ ($o\neq f$) for all other channels, one can solve
$P_f=T_f/(T_f+\sum_{o\neq f}T_o)$
for $T_f$ to be used e.g. for neutron induced cross section predictions (or also beta-delayed fission).
For applications to astrophysical question see e.g. \citep{Martinez-Pinedo.Mocelj.ea:2007,Petermann.Langanke.ea:2012,Erler.Langanke.ea:2012,Eichler.Arcones.ea:2015,Giuliani.ea:2018,Eichler.Sayar.ea:2019,Giuliani.Martinez.ea:2020,Kullmann.ea:2023,Holmbeck.Sprouse.ea:2023}.

\subsection{Level Densities}

The present status of the theoretical understanding of nuclear level densities
(one of the most important statistical nuclear properties), from empirical models to microscopic methods, has been given
in an excellent review \citep{Alhassid:2021}.
Many statistical model calculations use the level density description of
the back-shifted Fermi gas \citep{Gilbert.Cameron:1965}

\begin{align}
\rho(U,J,\pi) &= 1/2 f(U,J) \rho (U) \nonumber \\
\rho(U) &= {1 \over {\sqrt{2\pi} \sigma}} {\sqrt {\pi} \over {12a^{1/4}}}
{{\exp(2\sqrt{aU})} \over U^{5/4}} \label{eq:backshift}\\
f(U,J) &= { {2J+1} \over {2\sigma^2}} \exp(-J(J+1)/2\sigma^2) \nonumber \\
\sigma^2 = {\Theta_{rigid} \over \hbar^2} \sqrt{U \over a} &\ \ \ \ \
\Theta_{rigid} = {2 \over 5} m_uAR^2 \ \ \ \ \  U=E-\delta,\nonumber
\end{align}
\par\noindent
which assumes that positive and negative parities are evenly distributed
and that the spin dependence $f(U,J)$ is determined by the spin cut-off
parameter $\sigma$. The level density of a nucleus is therefore dependent
only on two parameters: the level density parameter $a$ and the backshift
$\delta$, which determines the energy of the first excited state. 
This backshift is related to the level spacing between the last occupied
and the first unoccupied
state in the single particle shell model and to the pairing gap, the energy
necessary to break up a proton or neutron pair. Of these two effects, the
pairing energy dominates except in the very near vicinity of closed shells.
Within this framework, the quality of level density
predictions depends on the reliability of systematic estimates of the level
density parameter
$a$ and the backshift $\delta$. Eq.(\ref{eq:backshift}) provides a valid functional
form for reproducing level densities down to an excitation energy of
$U=$1-2 MeV, but it diverges for $U=0$, i.e. $E=\delta$. This causes
problems if $\delta$ is a
positive backshift. Ericson plots, which show the number of excited
states as a function of excitation energy, reveal an almost linear
behavior for $\ln(N)=f(E)$ with an intercept at $E=E_0=\delta$ \citep{Ericson:1959}. 
This resulted in the ansatz $\rho(U)=\exp(U/T)/T$, commonly
called the constant temperature formula. The two formulations can be
combined for low and high excitation energies, with $E_0=\delta$ and $T$
being determined by a tangential fit to the Fermi gas formula. This description is 
usually called the composite (Gilbert-Cameron) formula.

The first compilation of $a$ and
$\delta$ was provided for a large number of nuclei by \cite{Gilbert.Cameron:1965}. They found that the backshift
$\delta$ is well reproduced by experimental pairing corrections.
Theoretical predictions result in $a/A\approx 1/15$ for infinite nuclear
matter, but the inclusion of surface and curvature effects of finite nuclei
enhances this value to 1/6 $-$ 1/8. While in principle $a$ can be dependent on temperature or 
excitation energy, such a dependence is
minimal for excitation energies up to 3$\times A$ (MeV) and thus the zero
temperature value for $a$ can be used up to these energies, covering
all values of interest within the
present context.
When comparing experimentally derived values of $a$ with such a simple formulation, one notices large deviations
due to shell effects (nuclei near closed shells have much smaller values for
$a$). An empirical correlation with experimental shell corrections $S(N,Z)$ was realized \citep{Gilbert.Cameron:1965}
\begin{equation}
{a/A} = c_0 + c_1 S(Z,N), \label{eq:gilcam}, 
\end{equation}
where $S(N,Z)$ is negative near closed shells. Many other functional
dependencies have been proposed and more microscopic treatments have been carried out 
\citep[see e.g.][]{Goriely.Hilaire.Girod:2012}. If one constructs level densities from single particle spectra,
it means that large care has to be invested in predicting correct single
particle spectra.

There have been a number of compilations for $a$ and $\delta$, or $T$
and $E_0$, based on purely experimental level densities, but predictions for unstable nuclei or 
those for which  experimental information is not available 
have to be based on theory, including the predictions for shell
and pairing corrections. For a review of early investigations see \cite{Cowan.Thielemann.Truran:1991},
utilizing the pairing corrections of a droplet nuclear mass model. Such an
average behavior changes, however, within one unit of magic nucleon numbers, where the
backshifts are often much larger. 

Deformed nuclei should in principle, be treated differently, as a
rotational band can be associated with each excited state. At high
excitation energies this treatment can, however, lead to double counting. 
Another option is to keep the same formalism as in Eq.(\ref{eq:gilcam}) but
performing an independent evaluation of the coefficients $c_0$ and $c_1$
for deformed nuclei, because only one
parameter set for $c_0$ and $c_1$, cannot achieve a significantly better
agreement with experimental level densities than found in \cite{Gilbert.Cameron:1965}
with a deviations up to a factor of 10. By dividing the nuclei into three classes (a: those
within three
units of magic nucleon numbers, b: other spherical nuclei, c: deformed
nuclei),
an improved agreement was obtained (maximum deviations of less than a factor
of 3 where experimental information at the neutron separation energy is available) \citep{Cowan.Thielemann.Truran:1991}.
However, this treatment was still very phenomenological and led \cite{Rauscher.Thielemann.Kratz:1997} to make use of
an improved approach, considering that the energy dependence
of the shell effects vanish at high excitation energies. Although, for astrophysical purposes
only energies close to the particle separation thresholds
have to be considered, an energy dependence can lead to
a considerable improvement of the global fit. This is especially
true for strongly bound nuclei close to magic numbers.
An excitation-energy dependent description has been
proposed for the level density parameter a \citep{Ignatyuk.ea:1980}
\begin{align}
a(U,Z,N)&=\tilde{a}(A)[1+C(Z,N) {f(U)\over U}] \nonumber \\
\tilde{a}&=\alpha A + \beta A^{2/3} \\
f(U)&=1-exp(-\gamma U). \nonumber
\end{align}
    
The values of the free parameters $a$, $b$, and $\gamma$ can be determined
by fitting to experimental level density data.
The shape of the function $f(U)$ permits the two extremes
(i) for small excitation energies the original form of Eq.(\ref{eq:gilcam})
is retained with $S(Z,N)$ being replaced by $C(Z,N)$ and (ii)
for high excitation energies $a/A$ approaches the continuum
value obtained for infinite nuclear matter \citep[for details see][]{Rauscher.Thielemann.Kratz:1997}.

The approach to the backshift $\delta$ could be improved further by utilizing the masses of neigboring nuclei
(from experiment of mass models if those are not available)
in order to obtain a more realistic pairing correction.
When obtaining the pairing correction directly from the masses
with 

\begin{align}
\delta &= {1 \over 2}[\Delta_n + \Delta_p]\nonumber\\
\Delta_n&= {1 \over 2} [-M(A-1,Z)+2 M(A,Z)-M(A+1,Z)] \label{eq:paring}\\
\Delta_p&= {1 \over 2} [-M(A-1,Z-1)+2 M(A,Z)-M(A+1,Z+1)]\nonumber
\end{align}

better fits to the total level density could be achieved with experimental level densities at
the neutron separation energy within a factor of 2 to 3, as utilized in the NON-SMOKER code. 
Parity-dependent level density descriptions (not using the factor 1/2 in Eq.(\ref{eq:backshift}) have also been utilized for tests in application to cross section predictions
\citep{Mocelj.ea:2007,Loens.Langanke.ea:2008} and are the default option in the code
SMARAGD.

\subsection{Width Fluctuation Corrections}
In addition to the ingredients required for Eq.(\ref{eq:HFmu}), like the
transmission coefficients for particles and photons and the level densities, 
width fluctuation corrections $W(j,o,J,\pi)$ have to be
employed. They define the correlation factors with which all
partial channels of incoming particle $j$ and outgoing particle $o$,
passing through excited state $(E,J,\pi)$, have to be multiplied.
This is due to the fact that the decay of the state is not fully
statistical, but some memory of the way of formation is retained and
influences the available decay choices. The major effect is elastic
scattering, the incoming particle can be immediately re-emitted before
the nucleus equilibrates. Once the particle is absorbed and not
re-emitted in the very first (pre-compound) step, the equilibration is
very likely. This corresponds to enhancing the elastic channel by a
factor $W_j$. The NON-SMOKER code uses the description of \cite{Tepel.Hofmann.Weidenmueller:1974},
leading to Eqs.(\ref{eq:wfc}) and (\ref{eq:wfc2}). In order to conserve the total cross
section, the individual transmission coefficients in the outgoing
channels have to be renormalized to $T_j^\prime $. The total 
cross section is proportional to $T_j$ and one obtains the condition 
$T_j$=$T_j^\prime (W_jT_j^\prime/T^\prime_{tot})+T^\prime_j(T^\prime_{tot}
-T^\prime_j)/T^\prime_{tot}$. This can be (almost) solved for $T^\prime_j$

\begin{equation}
T^\prime_j={T_j\over 1+ T^\prime_j(W_j-1)/T^\prime_{tot}}. \label{eq:wfc}
\end{equation}

This approach requires an iterative solution for $T^\prime$ (starting in the
first iteration with $T_j$ and $T_{tot}$), which converges fast.
The enhancement factor $W_j$ have to be known in order to apply
Eq.(\ref{eq:wfc}), a general expression 
in closed form is very complicated and computationally expensive to
employ. A fit to results from Monte Carlo calculations \citep{Tepel.Hofmann.Weidenmueller:1974}
leads to

\begin{equation}
W_j=1+{2\over 1+ T_j^{1/2}}. 
\label{eq:wfc2}
\end{equation}

For a general discussion of approximation methods \citep[see e.g.][]{Gadioli.Hodgson:1992}.
There exist further expressions for this renormalization \citep{Hofmann.ea:1980,Moldauer:1980}
which are utilized as options in the TALYS code. The NON-SMOKER code makes use of \cite{Tepel.Hofmann.Weidenmueller:1974} with
Eqs.(\ref{eq:wfc}) and (\ref{eq:wfc2}), which redefine the transmission
coefficients in Eq.(\ref{eq:HF*}) in such a manner that the total width is
redistributed by enhancing the elastic channel and weak channels over
the dominant one. While this is only an approximation to the
correct treatment, it could be shown that this treatment is quite adequate \cite{Thomas.Zirnbauer.Langanke:1986}. 

\subsection{Cross section applications}

Our discussion in the previous subsections reveals that one can
expect to obtain good agreement with experiment for particle and photon
transmission coefficients in comparison to experiments. On the other hand, level density predictions which are 
based on shell correction terms of
of microscopic-macroscopic mass models, show a statistical spread (a factor of 2
at the neutron separation energy). But all this applies to nuclei where experimental information is available, and 
can clearly be larger for nuclei far from stability. Neutron capture cross sections at 30 keV (utilized for many
astrophysical applications) of the NON-SMOKER code show agreement by less than a factor of 2, with
a few exceptions for nuclei located specifically at magic numbers. This
is a consequence of the still imperfect predictions of nuclear level
densities.

\begin{figure}[h!]
    \centering
    \includegraphics[width=0.6\textwidth]{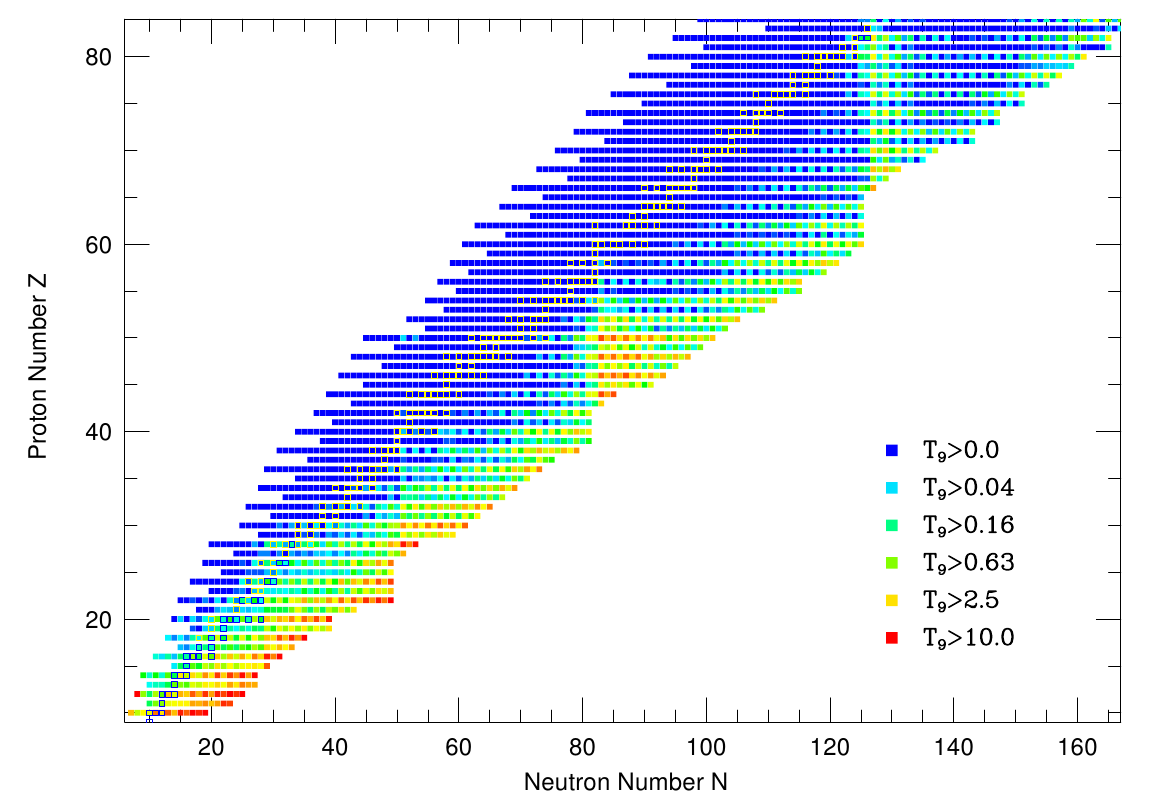}
    \caption{Required minimum temperatures for the applicability of the Hauser-Feshbach statistical model for neutron capture reactions, based on the request that about 10 resonances (levels) exist in the thermal energy window when utilizing a specific level density description \citep{Rauscher.Thielemann.Kratz:1997}. While in most explosive nucleosynthesis application such conditions are fulfilled, r-process applications in the range of 1-3$\times 10^9$K might be marginal close to the neutron-drip line just above neutron shell closures. These predictions were done by utilizing the microscopic-macroscopic FRDM model \citep{moeller95}, updates of this mass model and microscopic level density treatments might improve this situation. A similar plot for alpha-capture reactions \citep{Rauscher.Thielemann.Kratz:1997} leads to the conclusion that no temperature restrictions apply, because alpha-capture Q-values do not experience such a dependence of the distance from stability. For proton-capture reactions restrictions also apply close to the proton-drip line up to $Z\approx30$, requiring temperatures in excess of 2.5$\times 10^9$K \citep{Rauscher:2011,Rauscher:2020}. Image reproduced with permission of \cite{Rauscher.Thielemann.Kratz:1997}, copyright by APS.} 
    \label{fig:thermaln}
\end{figure}

This includes the question whether the statistical Hauser-Feshbach formalism can be applied to all nuclei, also far from stability. The statistical model assumes that one can utilize a transmission behavior averaged over resonances. This requires that in the Gamow window for charged particle reactions or in the energy window of a thermal neutron distribution a sufficient number of resonances is available in the compound nucleus. This energy window depends on the temperature in the plasma of reacting nuclei, because in astrophysical applications the reaction rate $\langle \sigma v\rangle$ depends on an integration of the cross section over a thermal distribution of reactants as shown in Eq.(\ref{sigmav}). Requiring about 10 levels in this thermal window leads to the following applicability criterion for neutron capture cross sections shown in Fig.\ref{fig:thermaln}. A major question is whether one would need to utilize direct capture predictions for the marginal regions in the nuclear chart, which include large uncertainties \citep{Mathews1983,Rauscher.ea:1998,Goriely:1998}, or a modification of
the statistical model \citep{Rauscher:2011}.

\begin{figure}[h!]
    \centering
    \includegraphics[width=0.8\textwidth]{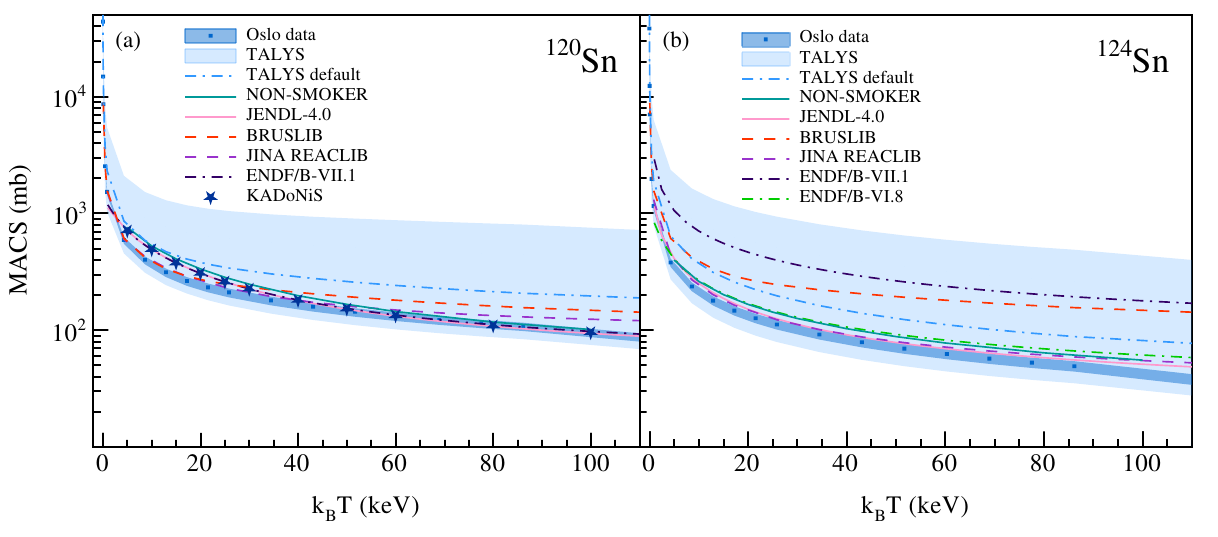}
    \caption{(Courtesy of M. Markova) Maxwellian averaged neutron capture cross section prediction for the Sn isotopes 120 and 124 as a function of the astrophysical environment temperature. Shown are the results based on experiments and the predictions of the NON-SMOKER as well as TALYS codes (for the latter indicates the spread, which results from the variety of options available).}
    \label{fig:markova2}
\end{figure}

Fig.\ref{fig:markova2} shows a comparison of Maxwellian averaged neutron capture cross sections for two Sn isotopes as a function of environment temperature. One can see that (still close to stability) the predictions of NON-SMOKER and the various options in the TALYS code provide a quite reasonable prediction within a factor of 2. The enhanced gamma-strength function due to low-lying strength of the PDR, M1 contributions and an upbend at lowest energies (see the subsection on $\gamma$-transmission coefficients), which can be included in the TALYS code, could, however, increase the difference between both codes for neutron-rich nuclei far from stability.

\begin{figure}[h!]
    \centering
    \includegraphics[width=0.5\textwidth]{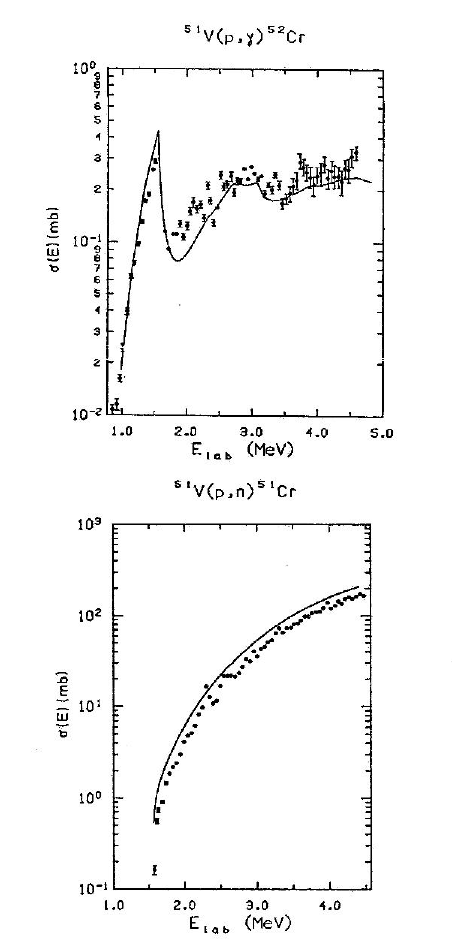}
    \caption{Cross section prediction for the reactions $^{51}$V($p,\gamma$) and $^{51}$V($p,n$), utilizing width fluctuation corrections \citep{Tepel.Hofmann.Weidenmueller:1974}. Without this correction the capture cross section would be enhanced by a factor of 4 beyond the neutron channel opening at 1.6 MeV. Image reproduced with permission from \cite{Cowan.Thielemann.Truran:1991}, copyright by Elsevier.}
    \label{fig:Wigner}
\end{figure}

At this point we also show some energy-dependent cross section
evaluations for charged particle reactions involving intermediate mass
nuclei which play an important role in silicon burning. 
Fig.\ref{fig:Wigner} indicates the importance of width fluctuation corrections when at a specific energy another reaction channel opens.

\begin{figure}[t!]
    \centering
    \includegraphics[width=0.8\textwidth]{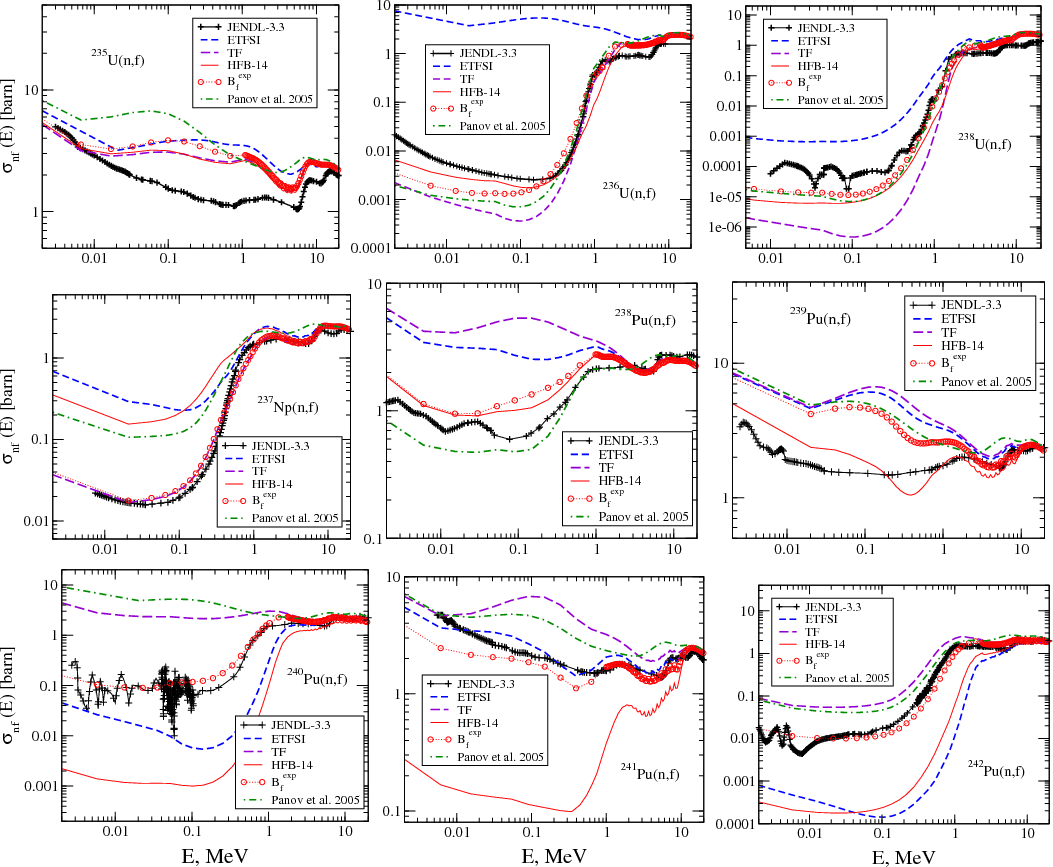}
    \caption{Neutron-induced fission cross sections predictions for a variety of actinide nuclei \citep{Panov.Korneev.ea:2010,Panov.Kolbe.ea:2005} in comparison to experiments from JENDL-3.3 \citep[crosses][]{Nakagawa.ea:2005}, averaged by the code JANIS \citep[black lines][]{Soppera.ea:2011}. It is clearly seen that the quality of theoretical fission barrier predictions \citep[ETFSI, TF, HFB-14,][]{Mamdouh.Pearson.ea:2001,Myers.Swiatecki:1999,Goriely.Hilaire.ea:2009} affects the results strongly. Image reproduced with permission from \cite{Panov.Korneev.ea:2010}, copyright by ESO.} 
    \label{fig:panov1}
\end{figure}

\begin{figure}[h!]
    \centering
    \includegraphics[width=0.7\textwidth]{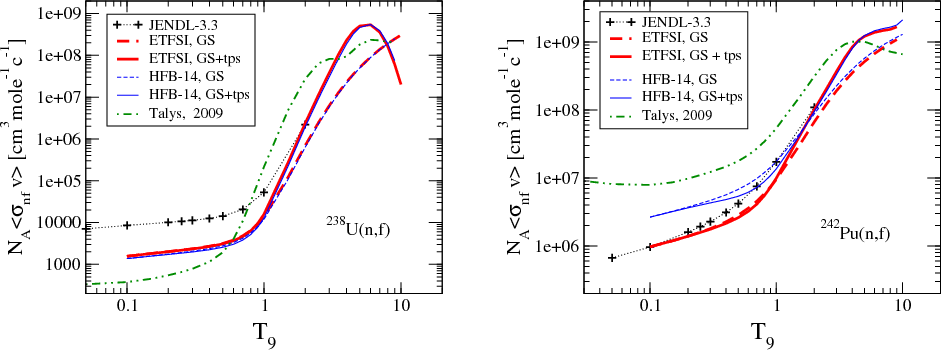}
    \caption{Similar to Fig.\ref{fig:panov1}, but for astrophysical neutron-induced fission reaction rates as a function of environment temperature \citep{Panov.Korneev.ea:2010}, utilizing either only the target ground state or thermally populated targets, which have an effect on the opening of the fission channel. TALYS (2009) stands for \cite{Goriely.Hilaire.ea:2009}, who include a thermal target population. Image reproduced with permission from \cite{Panov.Korneev.ea:2010}, copyright by ESO.}
    \label{fig:panov2}
\end{figure}

Finally we want to present also some results of fission cross section predictions \citep{Panov.Korneev.ea:2010}, which show that the procedure to predict these cross sections works quite well, but is highly sensitive to the quality of fission barrier predictions, when applied for unstable nuclei without experimental determinations of the fission barriers. Fig.\ref{fig:panov1} shows the prediction of neutron induced fission cross sections as a function of energy, (a) in comparison to experiments and (b) when applying different theoretical fission barrier predictions, Fig,\ref{fig:panov2} shows the astrophysical reaction rates as a function of environment temperature, also indicating when neutron energies surpass the fission barrier energies.

\section{Summary}
In the present Handbook chapter we examined the nuclear reactions in evolving stars, starting with hydrostatic burning phases of stellar evolution and continuing through explosive burning in various astrophysical environments, taking place in supernovae or explosive binary events. Important reactions were listed in Tables 1 through 7, going over to features being accompanied with the transition from dominating individual reactions to nuclear statistical equilibrium NSE in Figs. 1 through 4. Finally regions of the nuclear chart were identified, where nuclear reaction input is required for the different explosive nucleosynthesis processes. While the early phases of stellar burning, involving mostly light to intermediate mass nuclei, need to be tackled mostly by experimental approaches (see the chapter by M. Wiescher, J. deBoer \& R. Reifarth), for intermediate and heavy nuclei - with a high density of levels at the bombarding energy - statistical model (Hauser-Feshbach) approaches can be applied. This topic has been followed in detail, including the treatment of particle channels, gamma-transitions, fission, and level density predictions, also for nuclei far from stability. 

While our focus has been on charged-particle and neutron-induced reactions, we only mentioned the influence and importance of weak interactions in stellar environments, e.g. in late burning stages and in stellar explosions,
but want to refer here to the chapters by Suzuki \citep[and][]{Langanke.Martinez:2021} on electron capture reactions, and Rrapaj \& Reddy, Fuller \& Grohs, Wang \& Surman, Famiano et al., Fr\"ohlich, Wanajo, and Obergaulinger on the role of neutrino reactions in evolving stars and their explosive endpoint.

Finally it should be mentioned, that the recent white paper on nuclear astrophysics \citep{Schatz.ea:2022} is an enormous resource for presently ongoing experimental efforts from cross section measurements at astrophysical energies to investigations of nuclear properties with radioactive ion beam facilities far from stability. Experiments at underground facilities with the aim to measure at lowest possible energies, avoiding background noise are ongoing, e.g., at  LUNA\footnote{\url{https://luna.lngs.infn.it/index.php/new-about-us}}, CASPAR
\footnote{\url{https://caspar.nd.edu/}} and JUNA \citep{JUNA-Liu:2022}.

On the theoretical side
extended compilations for reaction rate predictions are provided in several publicly available database \footnote{\url{https://www.jinaweb.org/science-research/scientific-resources}},  \footnote{\url{https://nucastro.org}}, \footnote{\url{https://reaclib.jinaweb.org}}, \footnote{\url{https://www-nds.iaea.org}}, including results from the previously described codes NON-SMOKER and TALYS. All of these resources provide the input for the understanding of nuclear transmutations by reactions in evolving stars and their catastrophic endpoints \citep[for a more general overview about the astrophysical sites contributing to the cosmic abundances of all chemical elements and their role in the evolution of galaxies see, e.g.,][]{Arcones.Thielemann:2023}.

\subsection{Acknowledgements}
This article benefited from exchange and interactions within the European COST Action CA16117 Chemical Elements as Tracers of the Evolution of the Cosmos (ChETEC), and the International Research Network for Nuclear Astrophysics (IReNA). We also want to thank a large number of colleagues for communication and exchange related to the present overview, to name a few: Michael Wiescher, Ken'ichi Nomoto, Gabriel Martinez-Pinedo, Elena Litvinova, Maria Markova, Igor Panov, and Stephane Goriely.


\end{document}